\documentclass[reprint,
superscriptaddress,
 amsmath,amssymb,
 aps,
]{revtex4-1}

\usepackage{footmisc}
\usepackage{graphicx}
\usepackage{dcolumn}
\usepackage{bm}
\usepackage{xcolor}

\newcommand{\cconepi}{$\text{CC}1\pi^+$}

\begin{document}

\title{Measurement of the muon neutrino charged-current single $\pi^+$ production on hydrocarbon 
using the T2K off-axis near detector ND280 }


\newcommand{\INSTHD}{\affiliation{University Autonoma Madrid, Department of Theoretical Physics, 28049 Madrid, Spain}}
\newcommand{\INSTEE}{\affiliation{University of Bern, Albert Einstein Center for Fundamental Physics, Laboratory for High Energy Physics (LHEP), Bern, Switzerland}}
\newcommand{\INSTFE}{\affiliation{Boston University, Department of Physics, Boston, Massachusetts, U.S.A.}}
\newcommand{\INSTD}{\affiliation{University of British Columbia, Department of Physics and Astronomy, Vancouver, British Columbia, Canada}}
\newcommand{\INSTGA}{\affiliation{University of California, Irvine, Department of Physics and Astronomy, Irvine, California, U.S.A.}}
\newcommand{\INSTI}{\affiliation{IRFU, CEA Saclay, Gif-sur-Yvette, France}}
\newcommand{\INSTGB}{\affiliation{University of Colorado at Boulder, Department of Physics, Boulder, Colorado, U.S.A.}}
\newcommand{\INSTFG}{\affiliation{Colorado State University, Department of Physics, Fort Collins, Colorado, U.S.A.}}
\newcommand{\INSTFH}{\affiliation{Duke University, Department of Physics, Durham, North Carolina, U.S.A.}}
\newcommand{\INSTBA}{\affiliation{Ecole Polytechnique, IN2P3-CNRS, Laboratoire Leprince-Ringuet, Palaiseau, France }}
\newcommand{\INSTEF}{\affiliation{ETH Zurich, Institute for Particle Physics, Zurich, Switzerland}}
\newcommand{\INSTIE}{\affiliation{CERN European Organization for Nuclear Research, CH-1211 Genève 23, Switzerland}}
\newcommand{\INSTEG}{\affiliation{University of Geneva, Section de Physique, DPNC, Geneva, Switzerland}}
\newcommand{\INSTHJ}{\affiliation{University of Glasgow, School of Physics and Astronomy, Glasgow, United Kingdom}}
\newcommand{\INSTDG}{\affiliation{H. Niewodniczanski Institute of Nuclear Physics PAN, Cracow, Poland}}
\newcommand{\INSTCB}{\affiliation{High Energy Accelerator Research Organization (KEK), Tsukuba, Ibaraki, Japan}}
\newcommand{\INSTIB}{\affiliation{University of Houston, Department of Physics, Houston, Texas, U.S.A.}}
\newcommand{\INSTED}{\affiliation{Institut de Fisica d'Altes Energies (IFAE), The Barcelona Institute of Science and Technology, Campus UAB, Bellaterra (Barcelona) Spain}}
\newcommand{\INSTEC}{\affiliation{IFIC (CSIC \& University of Valencia), Valencia, Spain}}
\newcommand{\INSTHH}{\affiliation{Institute For Interdisciplinary Research in Science and Education (IFIRSE), ICISE, Quy Nhon, Vietnam}}
\newcommand{\INSTEI}{\affiliation{Imperial College London, Department of Physics, London, United Kingdom}}
\newcommand{\INSTGF}{\affiliation{INFN Sezione di Bari and Universit\`a e Politecnico di Bari, Dipartimento Interuniversitario di Fisica, Bari, Italy}}
\newcommand{\INSTBE}{\affiliation{INFN Sezione di Napoli and Universit\`a di Napoli, Dipartimento di Fisica, Napoli, Italy}}
\newcommand{\INSTBF}{\affiliation{INFN Sezione di Padova and Universit\`a di Padova, Dipartimento di Fisica, Padova, Italy}}
\newcommand{\INSTBD}{\affiliation{INFN Sezione di Roma and Universit\`a di Roma ``La Sapienza'', Roma, Italy}}
\newcommand{\INSTEB}{\affiliation{Institute for Nuclear Research of the Russian Academy of Sciences, Moscow, Russia}}
\newcommand{\INSTHI}{\affiliation{International Centre of Physics, Institute of Physics (IOP), Vietnam Academy of Science and Technology (VAST), 10 Dao Tan, Ba Dinh, Hanoi, Vietnam}}
\newcommand{\INSTHA}{\affiliation{Kavli Institute for the Physics and Mathematics of the Universe (WPI), The University of Tokyo Institutes for Advanced Study, University of Tokyo, Kashiwa, Chiba, Japan}}
\newcommand{\INSTID}{\affiliation{Keio University, Department of Physics, Kanagawa, Japan}}
\newcommand{\INSTIF}{\affiliation{King's College London, Department of Physics, Strand, London WC2R 2LS, United Kingdom}}
\newcommand{\INSTCC}{\affiliation{Kobe University, Kobe, Japan}}
\newcommand{\INSTCD}{\affiliation{Kyoto University, Department of Physics, Kyoto, Japan}}
\newcommand{\INSTEJ}{\affiliation{Lancaster University, Physics Department, Lancaster, United Kingdom}}
\newcommand{\INSTFC}{\affiliation{University of Liverpool, Department of Physics, Liverpool, United Kingdom}}
\newcommand{\INSTFI}{\affiliation{Louisiana State University, Department of Physics and Astronomy, Baton Rouge, Louisiana, U.S.A.}}
\newcommand{\INSTHB}{\affiliation{Michigan State University, Department of Physics and Astronomy,  East Lansing, Michigan, U.S.A.}}
\newcommand{\INSTCE}{\affiliation{Miyagi University of Education, Department of Physics, Sendai, Japan}}
\newcommand{\INSTDF}{\affiliation{National Centre for Nuclear Research, Warsaw, Poland}}
\newcommand{\INSTFJ}{\affiliation{State University of New York at Stony Brook, Department of Physics and Astronomy, Stony Brook, New York, U.S.A.}}
\newcommand{\INSTGJ}{\affiliation{Okayama University, Department of Physics, Okayama, Japan}}
\newcommand{\INSTCF}{\affiliation{Osaka City University, Department of Physics, Osaka, Japan}}
\newcommand{\INSTGG}{\affiliation{Oxford University, Department of Physics, Oxford, United Kingdom}}
\newcommand{\INSTGC}{\affiliation{University of Pittsburgh, Department of Physics and Astronomy, Pittsburgh, Pennsylvania, U.S.A.}}
\newcommand{\INSTFA}{\affiliation{Queen Mary University of London, School of Physics and Astronomy, London, United Kingdom}}
\newcommand{\INSTE}{\affiliation{University of Regina, Department of Physics, Regina, Saskatchewan, Canada}}
\newcommand{\INSTGD}{\affiliation{University of Rochester, Department of Physics and Astronomy, Rochester, New York, U.S.A.}}
\newcommand{\INSTHC}{\affiliation{Royal Holloway University of London, Department of Physics, Egham, Surrey, United Kingdom}}
\newcommand{\INSTBC}{\affiliation{RWTH Aachen University, III. Physikalisches Institut, Aachen, Germany}}
\newcommand{\INSTFB}{\affiliation{University of Sheffield, Department of Physics and Astronomy, Sheffield, United Kingdom}}
\newcommand{\INSTDI}{\affiliation{University of Silesia, Institute of Physics, Katowice, Poland}}
\newcommand{\INSTIA}{\affiliation{SLAC National Accelerator Laboratory, Stanford University, Menlo Park, California, USA}}
\newcommand{\INSTBB}{\affiliation{Sorbonne Universit\'e, Universit\'e Paris Diderot, CNRS/IN2P3, Laboratoire de Physique Nucl\'eaire et de Hautes Energies (LPNHE), Paris, France}}
\newcommand{\INSTEH}{\affiliation{STFC, Rutherford Appleton Laboratory, Harwell Oxford,  and  Daresbury Laboratory, Warrington, United Kingdom}}
\newcommand{\INSTCH}{\affiliation{University of Tokyo, Department of Physics, Tokyo, Japan}}
\newcommand{\INSTBJ}{\affiliation{University of Tokyo, Institute for Cosmic Ray Research, Kamioka Observatory, Kamioka, Japan}}
\newcommand{\INSTCG}{\affiliation{University of Tokyo, Institute for Cosmic Ray Research, Research Center for Cosmic Neutrinos, Kashiwa, Japan}}
\newcommand{\INSTHF}{\affiliation{Tokyo Institute of Technology, Department of Physics, Tokyo, Japan}}
\newcommand{\INSTGI}{\affiliation{Tokyo Metropolitan University, Department of Physics, Tokyo, Japan}}
\newcommand{\INSTHG}{\affiliation{Tokyo University of Science, Faculty of Science and Technology, Department of Physics, Noda, Chiba, Japan}}
\newcommand{\INSTF}{\affiliation{University of Toronto, Department of Physics, Toronto, Ontario, Canada}}
\newcommand{\INSTB}{\affiliation{TRIUMF, Vancouver, British Columbia, Canada}}
\newcommand{\INSTG}{\affiliation{University of Victoria, Department of Physics and Astronomy, Victoria, British Columbia, Canada}}
\newcommand{\INSTDJ}{\affiliation{University of Warsaw, Faculty of Physics, Warsaw, Poland}}
\newcommand{\INSTDH}{\affiliation{Warsaw University of Technology, Institute of Radioelectronics and Multimedia Technology, Warsaw, Poland}}
\newcommand{\INSTFD}{\affiliation{University of Warwick, Department of Physics, Coventry, United Kingdom}}
\newcommand{\INSTGH}{\affiliation{University of Winnipeg, Department of Physics, Winnipeg, Manitoba, Canada}}
\newcommand{\INSTEA}{\affiliation{Wroclaw University, Faculty of Physics and Astronomy, Wroclaw, Poland}}
\newcommand{\INSTHE}{\affiliation{Yokohama National University, Faculty of Engineering, Yokohama, Japan}}
\newcommand{\INSTH}{\affiliation{York University, Department of Physics and Astronomy, Toronto, Ontario, Canada}}

\INSTHD
\INSTEE
\INSTFE
\INSTD
\INSTGA
\INSTI
\INSTGB
\INSTFG
\INSTFH
\INSTBA
\INSTEF
\INSTIE
\INSTEG
\INSTHJ
\INSTDG
\INSTCB
\INSTIB
\INSTED
\INSTEC
\INSTHH
\INSTEI
\INSTGF
\INSTBE
\INSTBF
\INSTBD
\INSTEB
\INSTHI
\INSTHA
\INSTID
\INSTIF
\INSTCC
\INSTCD
\INSTEJ
\INSTFC
\INSTFI
\INSTHB
\INSTCE
\INSTDF
\INSTFJ
\INSTGJ
\INSTCF
\INSTGG
\INSTGC
\INSTFA
\INSTE
\INSTGD
\INSTHC
\INSTBC
\INSTFB
\INSTDI
\INSTIA
\INSTBB
\INSTEH
\INSTCH
\INSTBJ
\INSTCG
\INSTHF
\INSTGI
\INSTHG
\INSTF
\INSTB
\INSTG
\INSTDJ
\INSTDH
\INSTFD
\INSTGH
\INSTEA
\INSTHE
\INSTH

\author{K.\,Abe}\INSTBJ
\author{R.\,Akutsu}\INSTCG
\author{A.\,Ali}\INSTCD
\author{C.\,Alt}\INSTEF
\author{J.\,Amey}\INSTEI
\author{C.\,Andreopoulos}\INSTEH\INSTFC
\author{L.\,Anthony}\INSTFC
\author{M.\,Antonova}\INSTEC
\author{S.\,Aoki}\INSTCC
\author{A.\,Ariga}\INSTEE
% ...

\collaboration{The T2K Collaboration}\noaffiliation

\date{\today}

\begin{abstract}

We report the measurements of single and double differential cross section of muon neutrino charged-current interactions 
on carbon with a single positively charged pion in the final state at the T2K off-axis near detector 
using $5.56\times10^{20}$ protons on target.
The analysis uses data control samples for the background subtraction
and the cross section signal, defined as a single negatively charged 
muon and a single positively charged pion exiting from the target nucleus, 
is extracted using an unfolding method.
The model dependent cross section, integrated over the T2K off-axis
neutrino beam spectrum peaking at $0.6$~GeV, is measured to be
$\sigma = (11.76 \pm 0.44 \text{(stat)} \pm 2.39 \text{(syst)}) \times 10^{-40} \text{cm}^2$~$\text{nucleon}^{-1}$. Various differential cross sections are measured, including the first measurement of the Adler angles for single charged pion production in neutrino interactions with heavy nuclei target.

\end{abstract}

\maketitle

\section{Introduction\label{sec:intro}}

Precise knowledge of single charged pion production (\cconepi) induced by charged-current (CC) interactions of muon neutrinos with energy lower than a few GeV on nuclei is very relevant for current and upcoming neutrino oscillation experiments.
This process constitutes a background for the $\nu_\mu$ disappearance measurement when the charged pion is not observed.
In this energy range \cconepi has the largest neutrino interaction cross section after the CC quasi-elastic(CCQE) process. 
Single pion production is sensitive mainly to resonant processes but also to non-resonant contributions as well as coherent pion production. 
Moreover, in a nuclear target, there are multinucleon contributions and final state interactions to which the total and differential cross sections in pion kinematic variables are sensitive.
The correct modeling of these effects, interesting in its own right, is also a key challenge to the reduction of the systematic uncertainties in neutrino oscillation experiments.
A wide range of models exists and their validation requires
well-understood cross section measurements, both absolute and differential, and possibly on different nuclear targets.
To allow a meaningful comparison with different phenomenological models, the measured cross section data should be as independent as possible from the models themselves.

The first \cconepi cross section measurements are from decades-old hydrogen and deuterium bubble chamber experiments \cite{ANL,BNL}.
Despite the unsurpassed detector spatial resolution of bubble chambers, these results disagree by as much as 30\% due to large statistical uncertainties 
and poor modeling of the neutrino fluxes. 
Moreover, the uncertainties in nuclear effects make it difficult to extrapolate the cross sections to the heavier nuclei used as targets in modern neutrino experiments.
More recent measurements on different targets and energy ranges \cite{K2K-cc1pr,K2K-nc0pr,MiniBooNE-ratio,ANL-ratio} are
presented in the form of \cconepi to CCQE cross section ratios rather than absolute cross section measurement.

In recent years MiniBooNE~\cite{MiniBooNE-1pi},  MINER$\nu$A~\cite{MINERvA-1pi,MINERvA-1pi_1,MINERvA-1pi_2} and T2K~\cite{T2K1pi_water} reported absolute \cconepi cross sections, respectively in mineral oil, plastic scintillator and water, as a function of the relevant kinematic variables. These results show a significant disagreement, both in shape and in normalization~\cite{Betancourt:2018bpu,Sobczyk:2014xza}.
The difficulty of getting simultaneous agreement between all available low energy cross section data 
limits their effectiveness to constrain the uncertainty on cross section models and the corresponding systematic errors in neutrino oscillation experiments.

Since modern neutrino experiments use targets heavier than hydrogen and deuterium, it is not clear if the source of the discrepancy lies in the fundamental neutrino-nucleon cross section estimation or in the nuclear effects.
In a neutrino-nucleus interactions, the production of nucleons below the Fermi momentum is inhibited by the Pauli exclusion principle
and collective nuclear effects have to be considered.
Moreover, before leaving the target nucleus, 
interactions of the final state particles with the nuclear medium change their observed spectrum and composition. 
In particular, pion absorption and production change the event classification 
between \cconepi and other final states 
and in experimental measurements these effects cannot be  unfolded from the fundamental single nucleon cross section without relying on a specific model. Detailed understanding of the \cconepi interaction, such as the left-right asymmetry of the final-state hadron with respect to the lepton scattering plane, may help to constrain the absorbed pion background contribution to the CCQE-like neutrino interactions~\cite{Cai:2019jzk}.

Various models and implementations have been proposed \cite{Martini:2009,rein-sehgal,Salcedo:1987md,Hernandez:2007qq,Buss:2007ar,Praet:2008yn,Serot:2012rd,Ivanov:2015aya,Alam:2015gaa,Nakamura:2015rta,Hernandez:2013jka,Gonzalez-Jimenez:2016qqq,Sobczyk.:2012zj,Mosel:2017nzk,Kabirnezhad:2017jmf,Sobczyk:2018ghy} but since the size of the nuclear effects is large and there are discrepancies among models, it is important to provide experimental measurements that are as model-independent as possible. 
If the experimental signature is defined topologically by the particles leaving the target nucleus rather than the particles produced at the neutrino interaction vertex, the results can be compared with any specific model that combines nucleon-level cross section, nuclear effects and final state interactions.
This allows a thorough comparison with different predictions, reducing the modeling systematic uncertainties and easing the task of comparing different experimental results on the same target.
Robust experimental cross section data, and in particular \cconepi data, are needed to pin down which model, if any, gives the more accurate predictions and to assign a systematic uncertainty to it.

This paper describes the measurement of the \cconepi neutrino interaction cross section using the ND280 off-axis near detector in the T2K beam. The target material is plastic scintillator ($C_8H_8$) and the analysis selects charged current events with a negatively charged muon and a single positively charged pion, with no additional mesons but any number of additional nucleons. 

The paper is organized as follows. 
Section \ref{sec:expt} describes the key aspects of the neutrino beam and the ND280 detector used for this measurement. Section \ref{sec:analysis} describes the analysis strategy, the event selection and the candidates and control samples. The results are presented in Section \ref{sec:results} followed by conclusions in Section \ref{sec:conc}.

\section{Experimental setup\label{sec:expt}}

T2K is a long-baseline neutrino oscillation experiment located in Japan, whose goal is to make precise measurements of oscillation parameters via observation of muon (anti)neutrino disappearance and electron (anti)neutrino appearance \cite{t2k-nim}. 
A muon (anti)neutrino beam, produced in the J-PARC accelerator in Tokai, Japan, is directed at Super-Kamiokande, a large water Cherenkov detector located 295~km away near Kamioka. 
The beam is monitored by a set of near detectors that are additionally used for cross section measurements.

\subsection{Neutrino beam}
\label{beam}
The neutrino beam is initiated by collision of 30~GeV/c protons on a graphite target \cite{PhysRevD.87.012001}.
Resulting mesons (mainly pions) are collimated by three magnetic horns and enter a 96~m decay tunnel, where they decay into (anti-)neutrinos. 
Depending on the horns polarity, mesons of a desired sign are selected to produce a neutrino or antineutrino beam of high purity. 
For the data presented in this paper, the horns were operating in neutrino mode, focusing $\pi^+$ for a primarily $\nu_\mu$ beam. 

\begin{figure}[htbp]
 \includegraphics[width=0.49\textwidth]{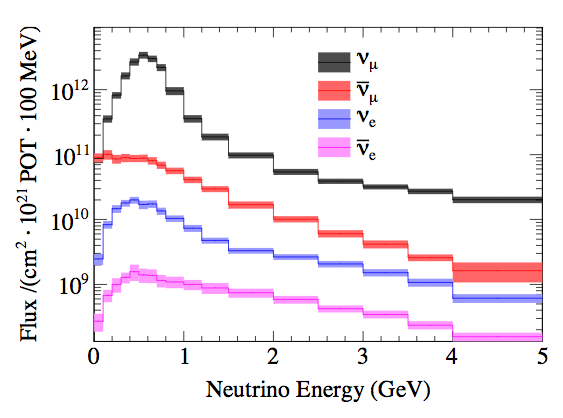}
 \caption{ND280 flux prediction with systematic error bars, for each neutrino flavor. \label{fig-t2kflux}}
 \end{figure}

The experiment uses an off-axis configuration, with detectors located away from the beam axis at an angle of 2.5$^{\circ}$, to get a narrow spectrum shape, optimised for oscillation studies. 
Beam stability and direction are monitored by a muon detector located at the end of the decay tunnel and by the INGRID near detector, which samples the neutrino beam on its central axis at approximately 280~m from the target. 
The predicted neutrino fluxes at the ND280 near detector, also located 280~m from the target, peaks at around 0.6~GeV and is shown in Figure~\ref{fig-t2kflux}. 
Muon neutrinos represent the largest fraction of the beam with 92.6\% of the total.  The remaining species are 6.2\% of $\bar{\nu}_\mu$, 1.1\% of $\nu_e$ and  0.1\% of $\bar{\nu}_e$.  

\subsection{Off-axis near detector ND280}
 
\begin{figure}[hbtp]
 \includegraphics[width=0.49\textwidth]{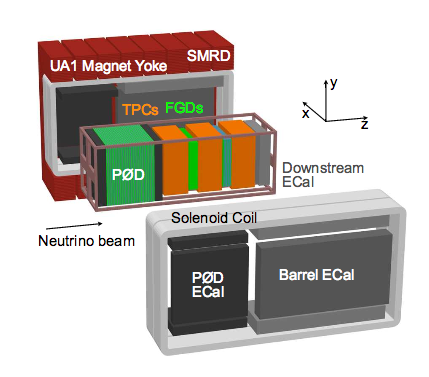}
 \caption{Schematic view of ND280 off-axis near detector. The Fine-Grained Detector (FGD), which provides the primary target mass for this measurement, has a cross sectional area of approximately 2~m x 2~m. The $\nu_{\mu}$ beam enters from the left of the figure.\label{fig-nd280}}
 \end{figure}

The off-axis near detector, ND280
is a magnetized particle tracking apparatus (see
Figure~\ref{fig-nd280}). 
Placed inside a magnet with a uniform dipole magnetic field of 0.2~T, it consists of a tracker and a $\pi^0$ detector (P0D)~\cite{t2k-p0d}, and is surrounded by electromagnetic calorimeters (ECals)~\cite{t2k-ecal} and Side Muon Range Detectors (SMRD)~\cite{Aoki:2012mf} . The tracker, located downstream of the P0D,
is made up of three gas Time Projection Chambers (TPCs)~\cite{t2k-tpc} interleaved with two Fine Grained Detectors (FGDs)~\cite{t2k-fgd}.

The FGDs are composed of finely segmented scintillator ($C_8H_8$) bars organized in layers. 
The orientation of the layers alternates between x and y directions almost perpendicular to the neutrino beam direction, allowing for precise reconstruction of the neutrino interaction vertex and track directions.
FGDs serve as the target for neutrino interactions in this analysis.
Their tracking capabilities provide tracks reconstruction down to a length of a few centimeters and evidence for additional activity around interaction vertices when tracks are too short to be reconstructed. The upstream FGD (FGD1) contains only active scintillator layers, while the downstream FGD (FGD2) incorporates also inactive water layers. 
To measure cross sections on $C_8H_8$, only neutrino interactions occurring in FGD1 are selected for this analysis.  
There are 30 scintillator layers in FGD1, with each layer containing 192 bars.  To reduce background from outside the FGD1 detector, a fiducial volume is defined by removing from the event selection events occurring inside any of the five bars at the edge of the detector in the transverse direction or in one of the two layers (one X and one Y projection) upstream of the neutrino beam direction.
The FGD1 fiducial volume has an elemental composition of  86.1\%  carbon  and  7.35\%  hydrogen  with  remaining contributions  from  oxygen  (3.70\%) and negligible quantities of other elements (Ti,  Si,  N). 

Three TPCs provide trajectory and energy loss information for tracks entering and exiting the FGDs, predominantly from muons and pions. Their capabilities allow for precise 3D track reconstruction, particle identification (PID) via the measurement of the ionization per unit length and determination of momentum
and charge by looking at curvature of tracks in the 0.2 T magnetic field. 

The ECals are sampling calorimeters consisting of layers of plastic scintillator separated by layers of lead.  
Alternating layers are aligned orthogonally to one another to provide three dimensional reconstruction of tracks and showers, both electromagnetic and
hadronic. 
The topological characteristics of the energy deposited in the ECal provide additional particle identification capability.

\section{Analysis description\label{sec:analysis}}

\subsection{Data samples definitions and observables }\label{sec:strategy}
In order to reduce the dependence from the modeling of final state particle re-interaction in the nuclear medium, the signal is defined in terms of the experimentally observable particles exiting the nucleus struck by the neutrino.  
The \cconepi final state is defined as one negatively charged muon, one and only one positively charged pion and any number of additional nucleons. Several additional control samples are selected 
to directly constrain with data the background subtraction. 
Restrictions are applied to the muon and pion kinematic in order to exclude phase space regions where the detection efficiency is low and the corresponding correction would introduce large model dependencies.

Seven differential cross section measurements are performed:
\begin{enumerate}
    \item {\it $d^2\sigma/dp_{\mu}d\cos\theta_{\mu}$}, where $p_{\mu}$ is the momentum of the muon and $\theta_{\mu}$ is the angle between the muon and the neutrino directions in the laboratory frame;
    \item {\it $d\sigma/dQ^2$}, where $Q^{2}$ is the reconstructed square of the 4-momentum transfer, defined from experimental observables in Eq.\ref{eq:Q**2};
    \item {\it $d\sigma/dp_{\pi}$}, where $p_{\pi}$ is the momentum of the pion in the laboratory frame;
    \item {\it $d\sigma/d\theta_{\pi \mu}$}, where $\theta_{\mu \pi}$ is the angle between the muon and the pion directions in the laboratory frame;
    \item {\it $d\sigma/d\theta_{\pi}$}, where ${\theta}_{\pi}$ is the angle between the pion and the neutrino directions in the laboratory frame;
    \item {\it $d\sigma/d\cos\theta_{Adler}$}, where $\cos\theta_{Adler}$ is defined as the polar angle in the Adler's coordinate system~\cite{Sanchez:2015yvw};
    \item {\it $d\sigma/d\phi_{Adler}$} where $\phi_{Adler}$ is defined as the azimuthal angle in the Adler's coordinate system~\cite{Sanchez:2015yvw}.
\end{enumerate}

The pion is identified either by a reconstructed TPC track or by the presence of a Michel electron detected in the FGD. In the latter case, the direction of the pion and its momentum is unknown.  For this reason the sub-sample of pions identified by the Michel electron is only used for the $d^2\sigma/dp_{\mu}d\cos\theta_{\mu}$.
The flux integrated differential cross sections are extracted using the D'Agostini unfolding method \cite{d'Agostini-unfolding} to correct for detector effects.

\subsection{Simulation}\label{sec:simulation}

Detector response, acceptance and efficiency are corrected using simulated Monte Carlo events to model the specific detector and beam configuration of each run with a sample ten times larger than the data statistics. 
The neutrino flux 
is predicted using simulations tuned to external measurements. Details of the beam simulation can be found in Ref.~\cite{PhysRevD.87.012001}. Interactions of protons in the graphite target and the resulting hadron production are simulated using the FLUKA 2011 package~\cite{Ferrari:2005zk,Battistoni:2007zzb}, weighted to match hadron production measurements~\cite{PhysRevC.85.035210,eichten,allaby,PhysRevC.77.015209,PhysRevC.84.034604,Abgrall:2015hmv}. The propagation and decay of those hadrons is performed in a GEANT3~\cite{GEANT3} simulation, which uses the GCALOR package~\cite{GCALOR} to model hadron re-interactions and decays outside the target. 
Uncertainties on the proton beam properties, horn current, hadron production model and overall neutrino beam alignment are taken into account to assess an energy-dependent systematic uncertainty on the neutrino flux. 
Flux tuning using NA61/SHINE data~\cite{PhysRevC.84.034604,PhysRevC.85.035210,Abgrall:2015hmv} reduces the uncertainty on the flux integrated overall normalization down to 8.5\%.
 
Neutrinos are propagated through the ND280 detector and their interactions with matter are simulated with the NEUT event generator. NEUT \cite{Hayato:2002sd, Hayato:2009} (version 5.1.4.2)
uses the Llewellyn-Smith CCQE neutrino-nucleon cross section formalism~\cite{llewellyn-smith} with the nuclear effects described by the Smith and Moniz~\cite{smith-moniz} relativistic Fermi gas (RFG) model. 
Dipole forms were used for both the axial and vector form factors. Tuning to Super-Kamiokande atmospheric data and K2K data led to set the nominal axial mass $M_\text{A}^\text{QE}$ to 1.21~GeV/c$^2$. 

The resonant pion production in NEUT is based on the Rein-Sehgal model~\cite{rein-sehgal}, taking into account 18 resonances with masses below 2~GeV/c$^2$ and their interference terms, with the axial mass $M_\text{A}^\text{RES} = 1.21$~GeV/c$^2$. 

Neutral Current (NC) and Charged Current (CC) coherent pion production is simulated using the Rein-Sehgal model in Ref.~\cite{rein-sehgal-coha}. The CC coherent pion production includes the PCAC (Partially Conserved Axial vector Current) and lepton mass corrections~\cite{rein-sehgal-cohb}.

DIS (Deep Inelastic Scattering) processes are simulated using the GRV98~\cite{Gluck:1998xa} parton distribution with low-Q$^2$ corrections by the Bodek and Yang model~\cite{Bodek:2003wd}.

Secondary interactions of pions inside the nucleus, so-called final state interactions (FSI) are simulated using an intranuclear cascade model based on the method described in Ref.~\cite{Salcedo:1987md}, tuned to external $\pi$-$^{12}$C data.

The GENIE \cite{Andreopoulos:2009rq} (version 2.6.4) neutrino generator is used as an alternative simulation to test the dependence of the analyses on the assumed signal and background models. 
Among other differences, GENIE uses different values of $M_\text{A}^\text{QE}$=0.99/c$^2$~GeV~\cite{Kuzmin:2007kr} and $M_\text{A}^\text{RES}$=1.12~GeV/c$^2$~\cite{Kuzmin:2006dh}. We did not observe any significant variation of the results using this alternative event generator. 

The simulated final state particles are then propagated through the detector material using GEANT4~\cite{GEANT4}.

\subsection{Event selection}\label{sec:preselection}

The analysis presented here uses data from the three T2K run periods between November 2010 and May 2013, where T2K was operating in neutrino mode. In total $5.56 \times 10^{20}$ protons on target (POT) are used, corresponding to all good quality data, with each sub-detector working optimally.

Events with the highest momentum track consistent with a negatively charged particle passing the TPC track quality selection criteria and matched with a track originating in the upstream FGD are selected as muon neutrino interactions candidates. The energy deposition measured in the TPC is required to be compatible with the energy loss of a muon-like, minimum ionizing particle. Further selection criterion are applied to remove events where the interactions occur outside the FGD fiducial volume. Further details on the $\nu_\mu$ CC inclusive selection can be found in~\cite{Abe:2015awa}. 

To further select \cconepi events, the presence of one and only one pion of positive charge is required.  The pion is identified by a positively charged TPC track with an energy deposition compatible with a pion or by the presence of a Michel electron, tagged as a time delayed energy deposition in the upstream FGD fiducial volume. The event is rejected if additional pions, either charged or neutrals, or photons are identified in the event either by looking at TPC tracks or electromagnetic showers in ECal.

Table~\ref{tab:effbycut_CC1Pion} shows the data and Monte Carlo reduction and the fraction of events that survive each selection criteria with respect to the previous one. Monte Carlo and Data survival fractions are similar after the quality, fiducial and backward tracks removal selection criterion are applied. These selection cuts eliminate events outside of the detector fiducial volume that are not properly simulated in our event generation. The composition of the selected sample according to the $\pi^+$ selection criteria is shown in Table~\ref{tab:SelComposition_CC1Pion}. The data sample has slightly more $\pi^+$'s events selected with the Michel electron criteria but still compatible within 1$\sigma$ statistical error. 

The MC events shown in Table~\ref{tab:effbycut_CC1Pion} and \ref{tab:SelComposition_CC1Pion} are bare predictions, they are not corrected by several effects such as the detection efficiency and the re-weight of the event generator probabilities. The correction is applied later in the analysis leading to a modification of the reported final cross-section.  

\begin{table}[htbp]
\caption{Number of events selected after each selection criteria. Monte Carlo events (NEUT) are normalized to the data POT. In parenthesis the fraction of events surviving each selection step with respect to the previous one is shown. }
\label{tab:effbycut_CC1Pion}
\begin{center} 
\begin{tabular}{|l | c | c|c|}
\hline
Selection criteria        &  Data events          &  MC events \\
\hline
Total multiplicity   & 1,927,791      &   1,041,707.5\\
Quality and Fiducial & 47,900 (24.4\%)& 35,550.2  (34.1\%) \\ 
Backward tracks  & 34,762 (74\%) & 28,545.2  (80\%) \\ 
Upstream veto & 33,660 (97\%) & 27,827.3  (97\%) \\
Muon PID   & 24,378 (72\%) & 20,012.3  (72\%) \\ 
One pion   & 2,739  (11\%) &  2588.1 (13\%) \\
\hline
\end{tabular}
\end{center} 
\end{table}

\begin{table}[htbp]
\caption{ Composition of the CC$1\pi^+$ selection according to the $\pi^+$ selection criteria. NEUT MC is normalised to the data POT.  The fractional errors are computed varying each sample independently according to a Poisson distribution. }
\label{tab:SelComposition_CC1Pion}
\begin{center}
\begin{tabular}{|c | c | c|c|}
\hline
$\pi^+$ selection &  Data events          &  NEUT MC events \\
criteria & & \\
\hline
TPC track   & 1563  (57.06$\pm$0.95\%)  & 1503.9   (58.11$\pm$0.31\%)  \\
Michel electron & 1176  (42.94$\pm$0.95\%) & 1084.2  (41.89$\pm$0.31\%)  \\ 
\hline
\end{tabular}
\end{center} 
\end{table}

\subsection{Selected sample composition}
\label{sec:CCsplited}

Table~\ref{tab:purity_CCsplited} shows the composition of the selected \cconepi sample with respect to the true topology according to the NEUT Monte Carlo. The topologies are event classifications based on the number of pions leaving the nucleus: 1 $\mu^-$ and 0 pions (CC0$\pi$), 1 $\mu^-$ and a single $\pi^+$ (CC$1\pi^+$), one $\mu^-$ plus a $\pi^-$, a $\pi^0$ or more than one pion (CCOther), 0 $\mu^-$ (Background) or events produced outside the fiducial volume (OOFV). Table~\ref{tab:purity_CCsplited} shows also the  compositions of the full selected sample (second column) and of the two sub-samples where the pion is reconstructed in the TPC (third column) or identified by the Michel electron (fourth column). The largest contamination in the final sample comes from multi-pions interaction where the additional pions are absorbed in subsequent interactions with the detector material or simply not reconstructed. The main neutrino interaction process at the nucleon level contributing to the \cconepi sample as predicted by the event generation is the pion resonant production (61.5\%).

\begin{table}[htbp]
\caption{Composition of the \cconepi sample with respect to the true topologies for the full sample (second column), the sub-sample in which the pion is reconstructed in the TPC (third column) and the sub-sample in which the pion is identified by the presence of a Michel electron (fourth column). The ``Background'' component contains anti-neutrino, electron neutrino and neutral current events. Out of fiducial volume (OOFV) are interactions generated outside the FGD1 fiducial volume.}
\label{tab:purity_CCsplited}
\begin{center} 
\begin{tabular}{|l | c | c| c| r|}
\hline
Component & Full sample & $\pi^+$ TPC & Michel electron \\
\hline
CC0$\pi$    & 5.00   \% &   4.1 \% & 6.3 \%  \\
CC$1\pi^+$ & 61.5 \% &  61.1 \% & 62.0 \% \\
CCOther    & 22.0   \% &  24.7 \% & 17.5 \% \\
Background & 6.2  \% &   7.9 \% & 3.3  \% \\
OOFV       & 5.4   \% &   2.2 \% & 10.8 \% \\
\hline
\end{tabular}
\end{center} 
\end{table}

\subsection{Kinematic observables}

This section discusses the distributions of the reconstructed pion and muon basic kinematic variables for the selected sample.
Data are compared with the expectations of the NEUT and GENIE Monte Carlo generators in terms of the topologies introduced in Section~\ref{sec:CCsplited}.

\begin{figure*}[htb]
\centering 
  \includegraphics[width=.4\linewidth]{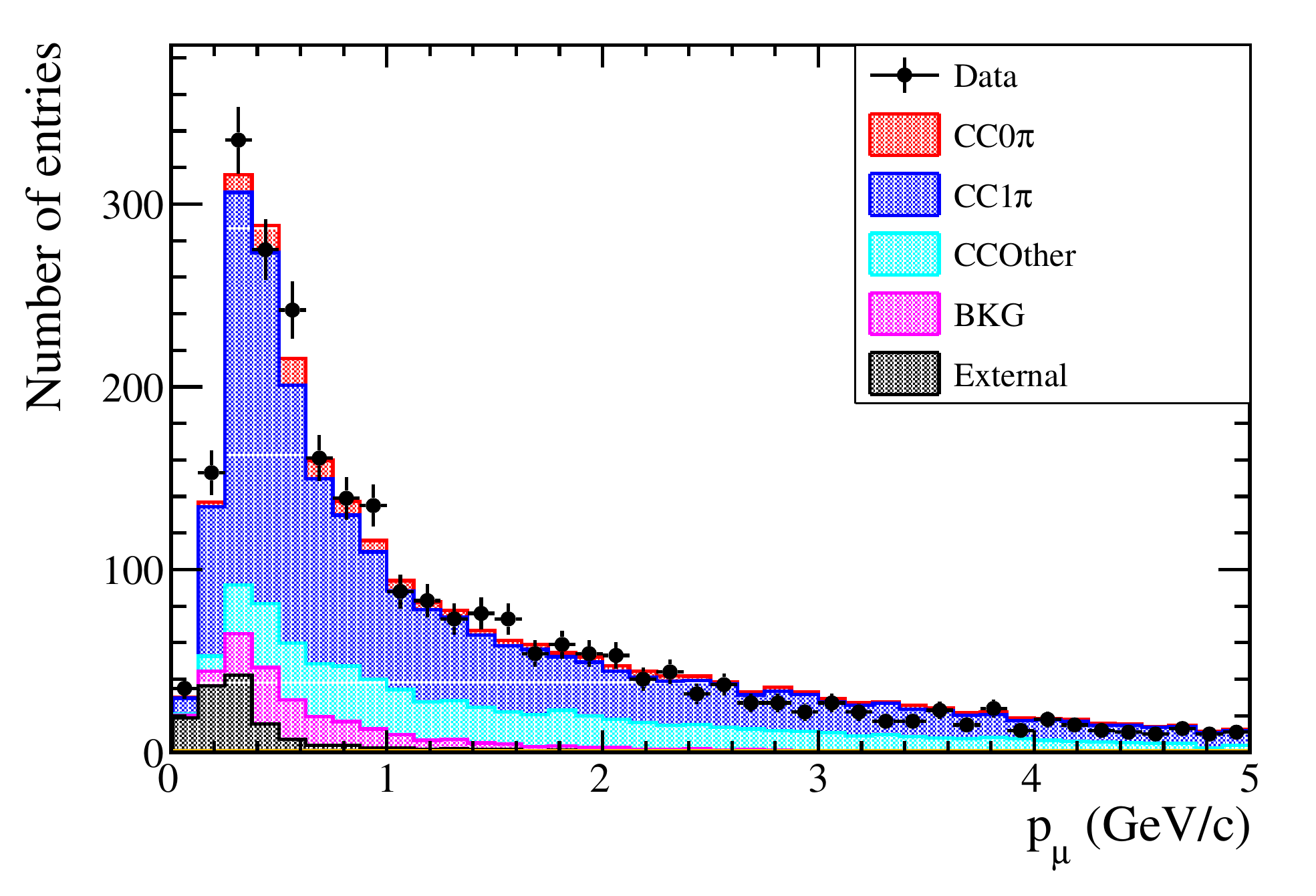}
  \includegraphics[width=.4\linewidth]{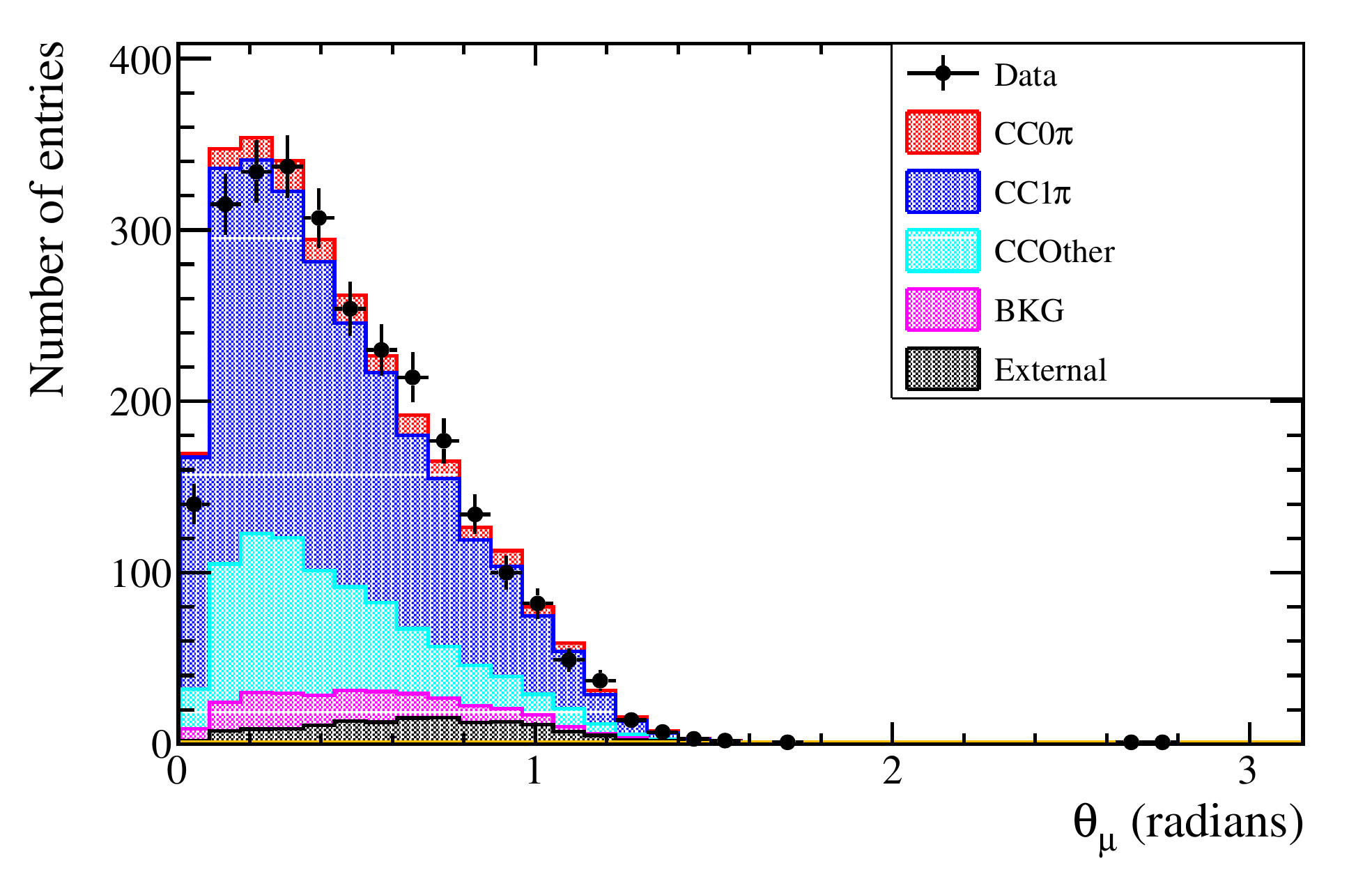}
  \includegraphics[width=.4\linewidth]{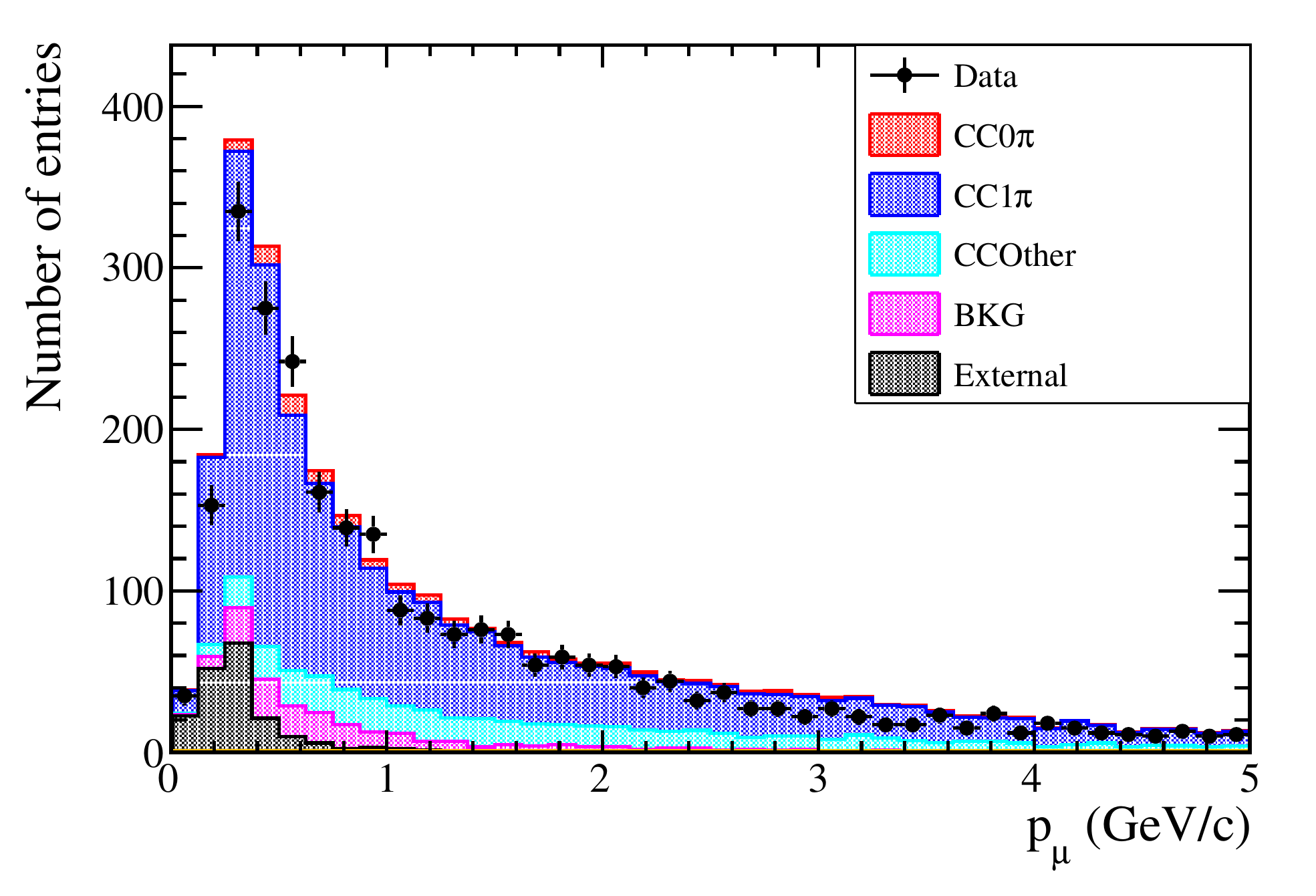}
  \includegraphics[width=.4\linewidth]{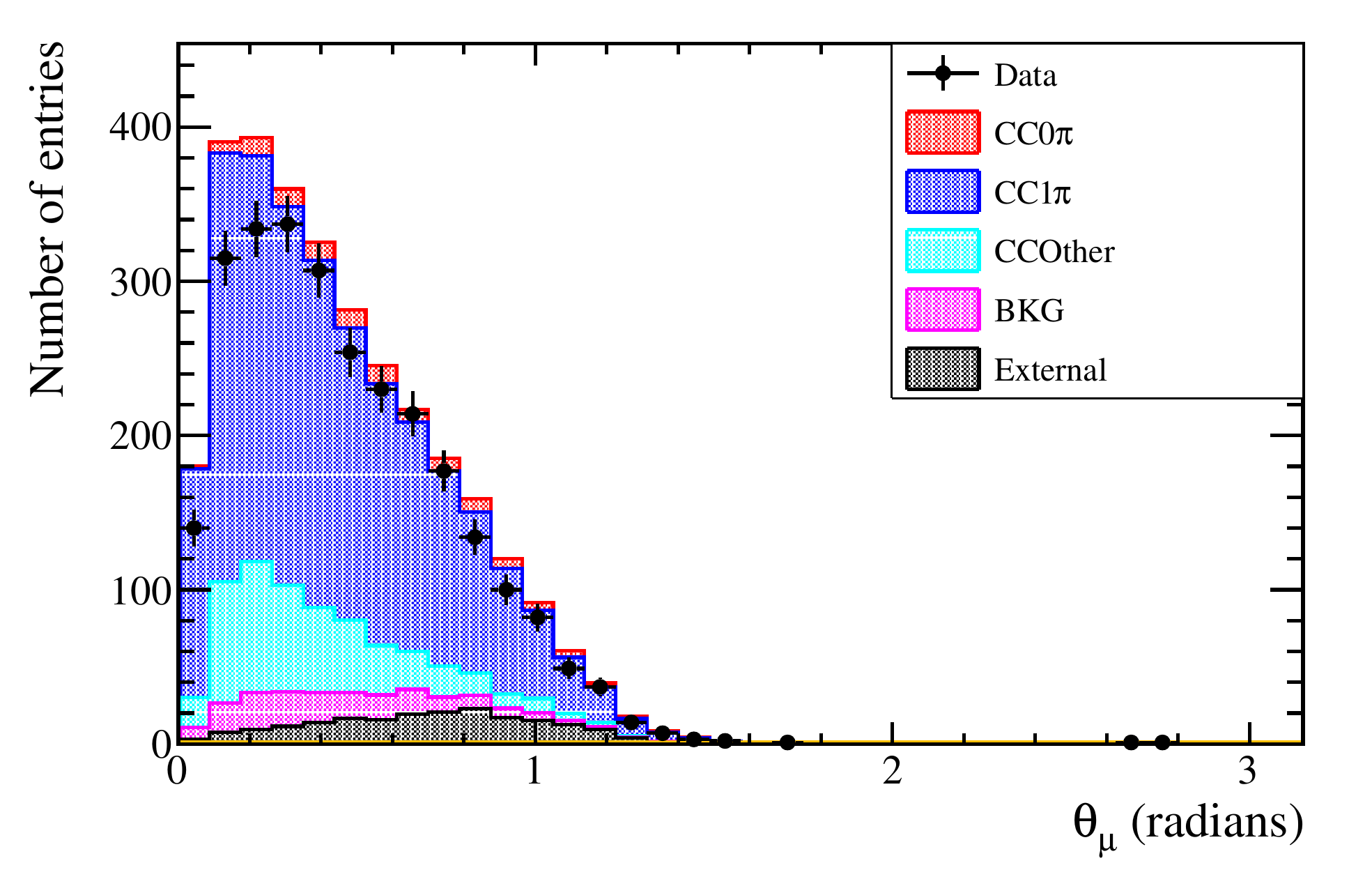}
  \caption{Muon momentum distribution (left) and muon angle (right)
    for the selected \cconepi sample. Data are compared with NEUT 5.1.4.2 (upper plots) and GENIE 2.6.4 (lower plots).}
\label{fig:MuMom}
\end{figure*}

Figure~\ref{fig:MuMom} shows the distributions of muon momentum (left plots) and angle (right plots) for the selected \cconepi events compared with NEUT (upper plots) and GENIE (lower plots) Monte Carlo expectations.

\begin{figure*}[htb]
\centering
\includegraphics[width=.4\linewidth]{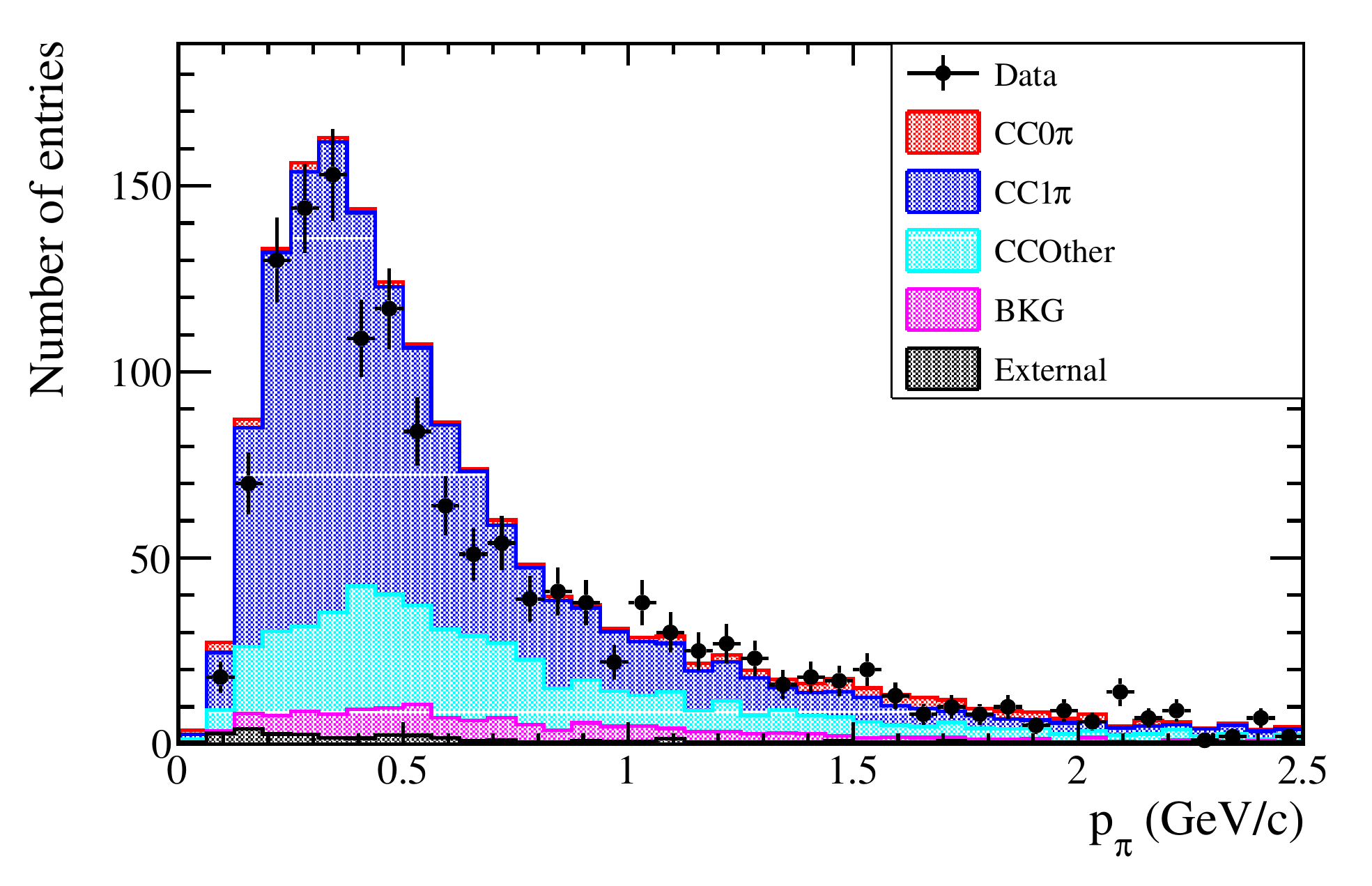}
\includegraphics[width=.4\linewidth]{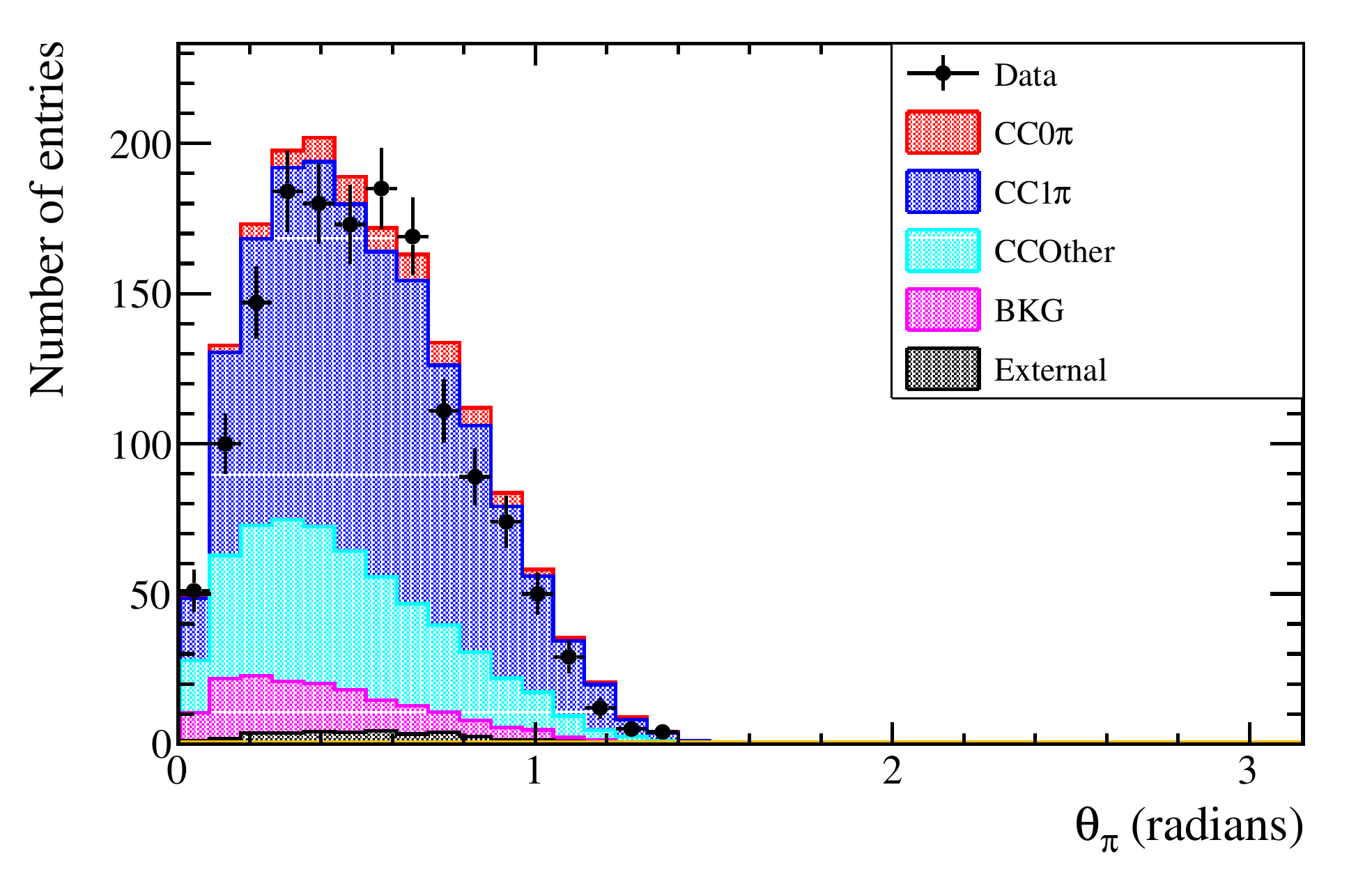}
\includegraphics[width=.4\linewidth]{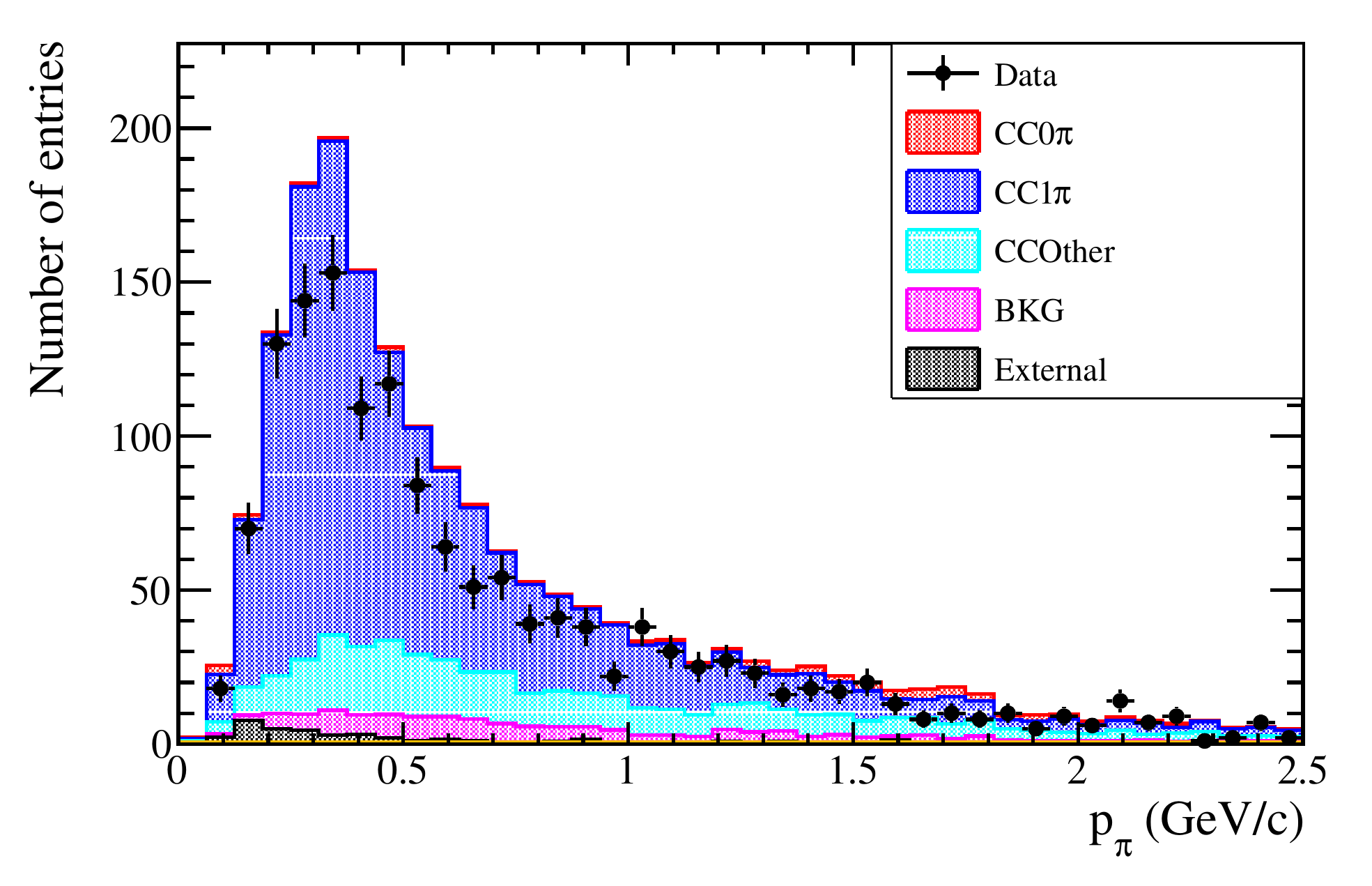}
\includegraphics[width=.4\linewidth]{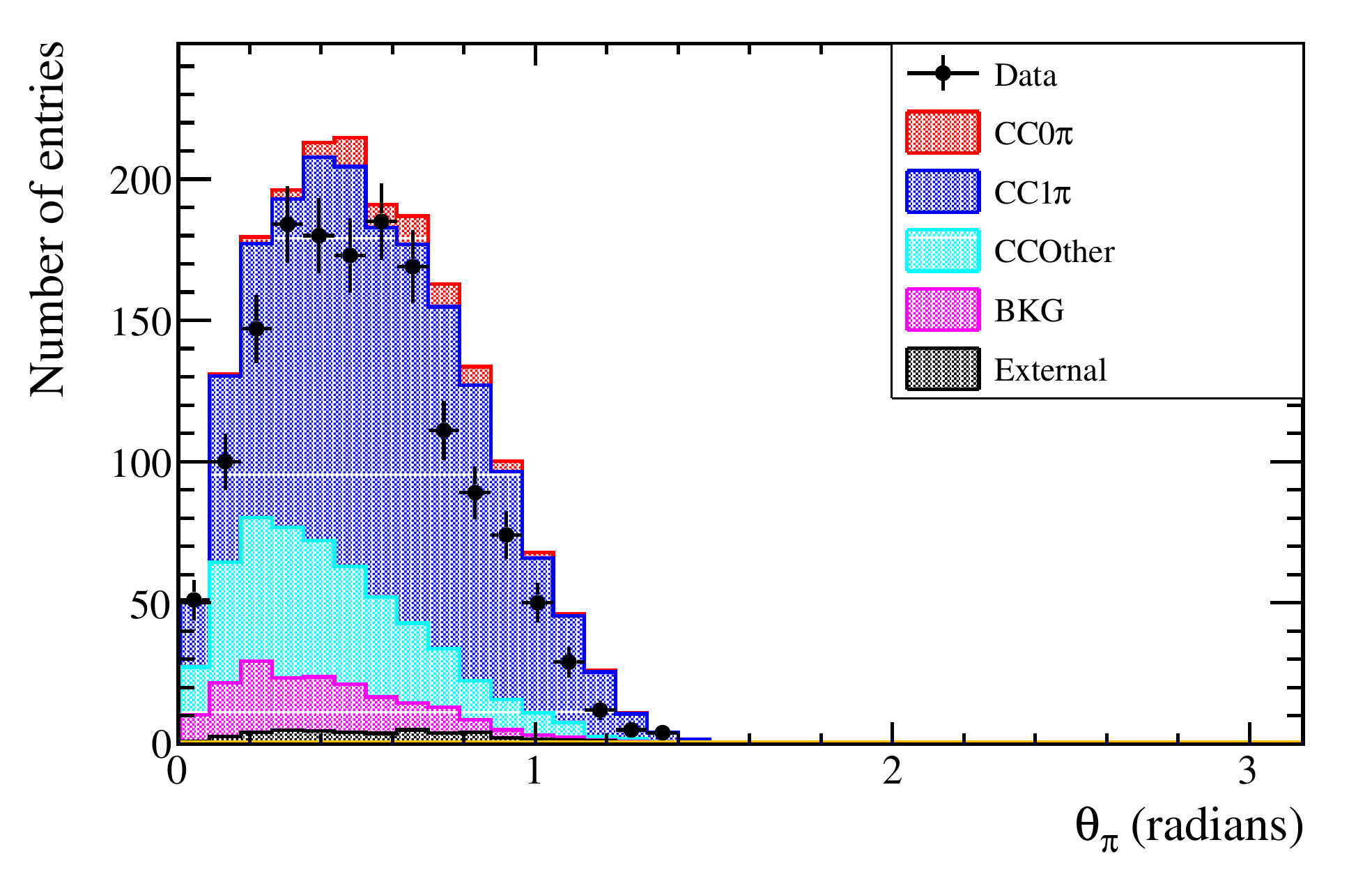}

\caption{Pion angle distribution (right) and pion momentum (left) for the \cconepi sub-sample of events with pion reconstructed in the TPC. Data are compared with NEUT 5.1.4.2 (upper plots) and GENIE 2.6.4 (lower plots). }
\label{fig:PiCos_Theta_mu_pi_Rec}
\label{fig:PiMom}
\end{figure*}

 The pion momentum distributions, see Figure~\ref{fig:PiMom}, are shown for the sub-sample of events where the pion is reconstructed in the TPC (left plots).  Data are compared with NEUT 5.1.4.2 (upper plots) and GENIE 2.6.4 (lower plots). Similarly, Figure~\ref{fig:PiCos_Theta_mu_pi_Rec} (right plots) shows the distribution of the pion angle with respect to the neutrino beam direction for the sub-sample with the pion direction reconstructed in the TPC. NEUT predictions before background subtraction show better agreement with the data than GENIE that predicts slightly more events. 

\subsubsection{Event kinematic observables}

\begin{figure*}[htp]
\centering 
  \includegraphics[width=.4\linewidth]{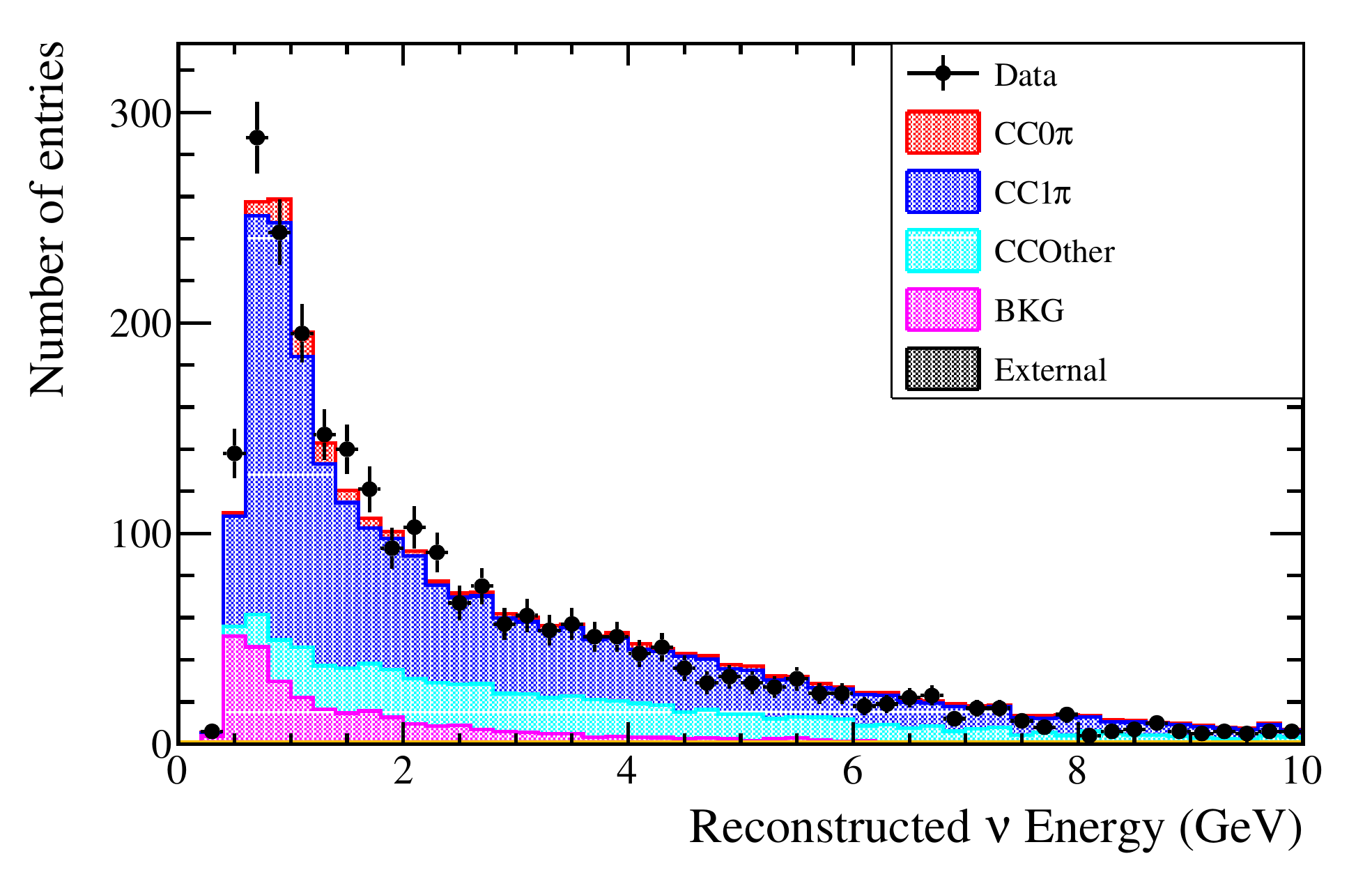}
 \includegraphics[width=.4\linewidth]{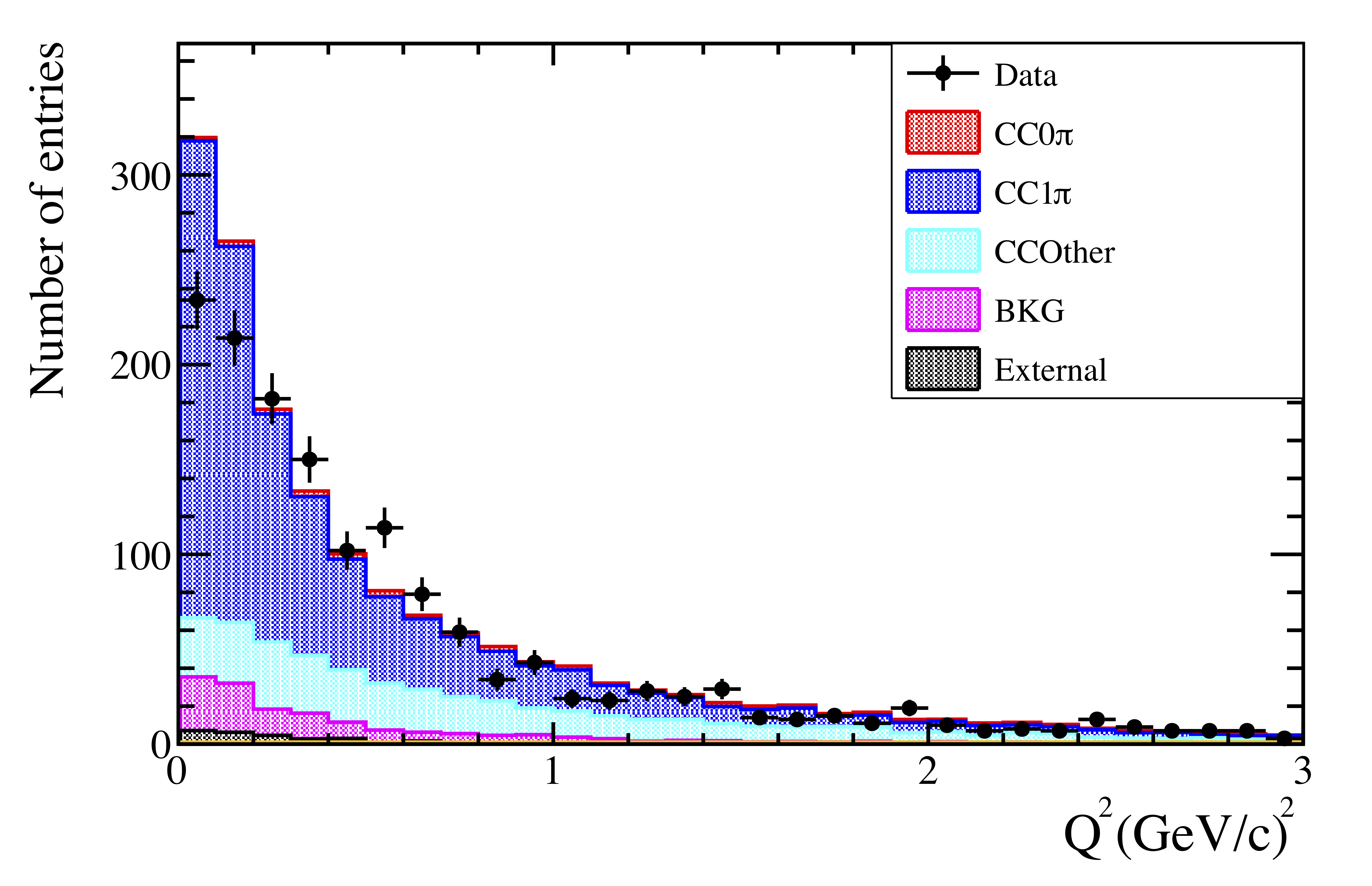}
  \caption{Reconstructed neutrino energy (left),
    4-momentum transfer squared of the interaction (right)
    for the \cconepi sub-sample of events with pion reconstructed in the TPC. Data are compared with NEUT 5.1.4.2.}
\label{fig:EnuExtended_Rec}
\label{fig:Q2Extended_Rec}
\label{fig:W_planar_Rec}
\end{figure*}

Kinematic variables like the neutrino energy and the momentum transfer are reconstructed from the muon and pion kinematics under the assumption that the nucleon struck by the neutrino is at rest, bound to the target nucleus by an energy $E_{bind}$(25 $MeV/c$), and the final state contains, beside the pion and the muon, a single undetected proton. The neutrino energy is reconstructed using the equation 

\begin{equation}\label{eq:enu}
E_{\nu}= \frac{m_p^2 - (m_p - E_{bind} - E_{\mu} - E_{\pi})^2 + |\overrightarrow{p}_{\mu} + \overrightarrow{p}_{\pi}|^2}{2(m_p - E_{bind} - E_{\mu} - E_{\pi} + \overrightarrow{d}_{\nu}\cdot(\overrightarrow{p}_{\mu} + \overrightarrow{p}_{\pi}))}.
\end{equation}

 where $m_p$ stands for the proton mass, $\overrightarrow{d}_{\nu}$ is the predicted neutrino direction and $(\overrightarrow{p}_{\mu,\pi},E_{\mu,\pi})$ are the reconstructed muon and pion four-momenta.  Figure~\ref{fig:EnuExtended_Rec} (left) shows the reconstructed neutrino energy distribution for the \cconepi events where the pion is reconstructed in the TPC.
 Neutrino direction ($\overrightarrow{d}_{\nu}$) is fixed along the neutrino flux thrust, although the Monte Carlo simulation includes an accurate description of its angular dispersion.

The 4-momentum transfer is defined as:
\begin{equation}\label{eq:Q**2}
  Q^2=-q^2=(p_{\mu}-p_{\nu})^2
\end{equation}
where $p_{\mu}$ and $p_{\nu}$ are the 4-momentum vectors of the muon and neutrino respectively. 
Figure~\ref{fig:Q2Extended_Rec} shows the $Q^2$ distribution for the candidate \cconepi events.

\subsubsection{Adler's angles } 

\begin{figure*}[htbp]
\centering
\includegraphics[width=0.99\linewidth]{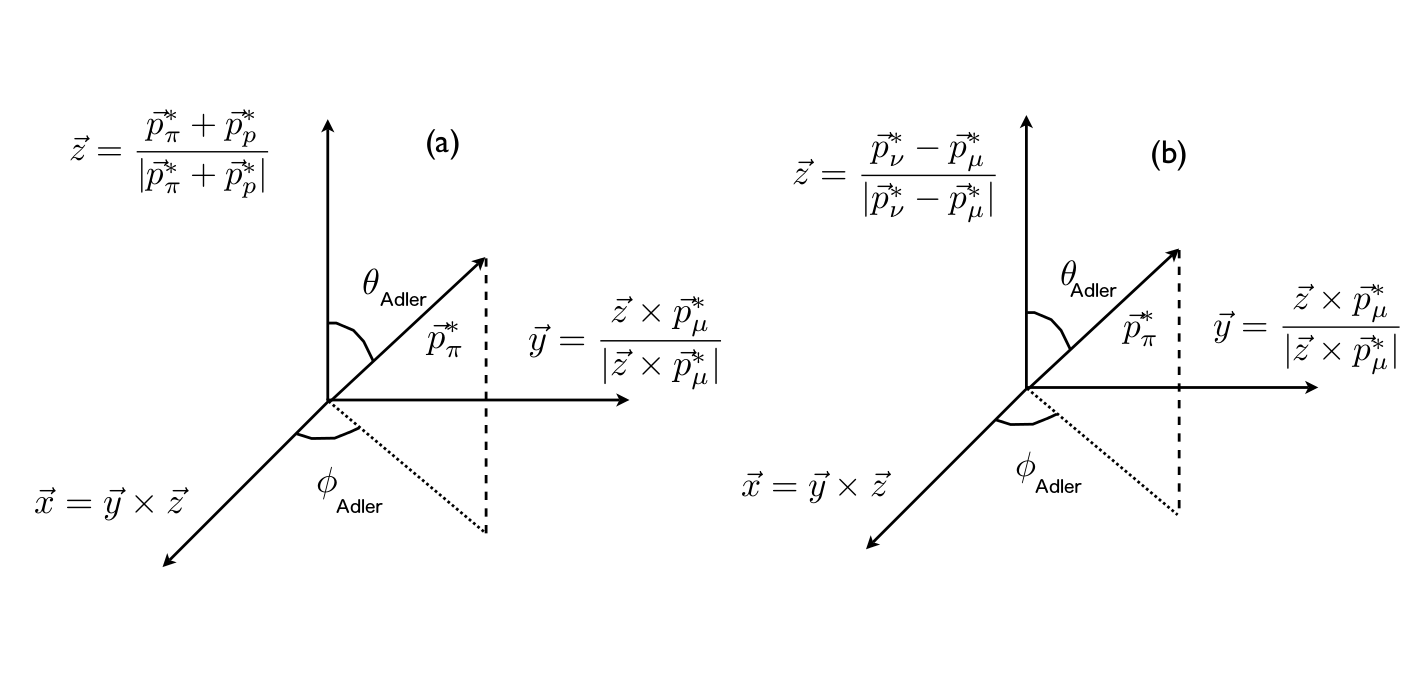}
\caption{Azimuthal and polar angles of the pion in the Adler reference system (left).
The Adler reference system computed using experimentally accessible observables (right).}
\label{fig:PolarAngles}
\end{figure*}

The angles, $\theta_{Adler}$ and $\phi_{Adler}$, define the direction of the pion in the Adler's system.  The Adler reference system is the 
$p\pi^+$ rest frame as shown in Figure~\ref{fig:PolarAngles} (left) where $p_{\mu}^*$, $p_{\pi}^*$ and $p_{p}^*$ correspond to the muon, pion and final state nucleon momentum.  
The angles $\theta_{Adler}$ and $\phi_{Adler}$ are sensitive respectively to the longitudinal and transverse polarization of the $p\pi^+$ final state for interactions mediated by the $\Delta^{+}$, $\Delta^{++}$ and non resonant contributions.
The experimental definition of the Adler system needs to be changed in terms of lepton and pion observables since the final state nucleon is not usually detected\cite{Sanchez:2015yvw}. The Adler rest frame and the angles $\theta_{Adler}$ and $\phi_{Adler}$ are redefined as shown in Figure~\ref{fig:PolarAngles} (right), where the neutrino direction is assumed known and the neutrino energy is reconstructed from Eq.~(\ref{eq:enu}). 
It has been shown \cite{Sanchez:2015yvw} that with this experimental definition the information of the original Adler angles is reasonably maintained when the neutrino interacts with light nuclei despite the need to determine the incoming neutrino energy from the lepton and pion observables and the effects of FSI in the target nucleus.

\begin{figure*}[htp]
\centering 
  \includegraphics[width=.4\linewidth]{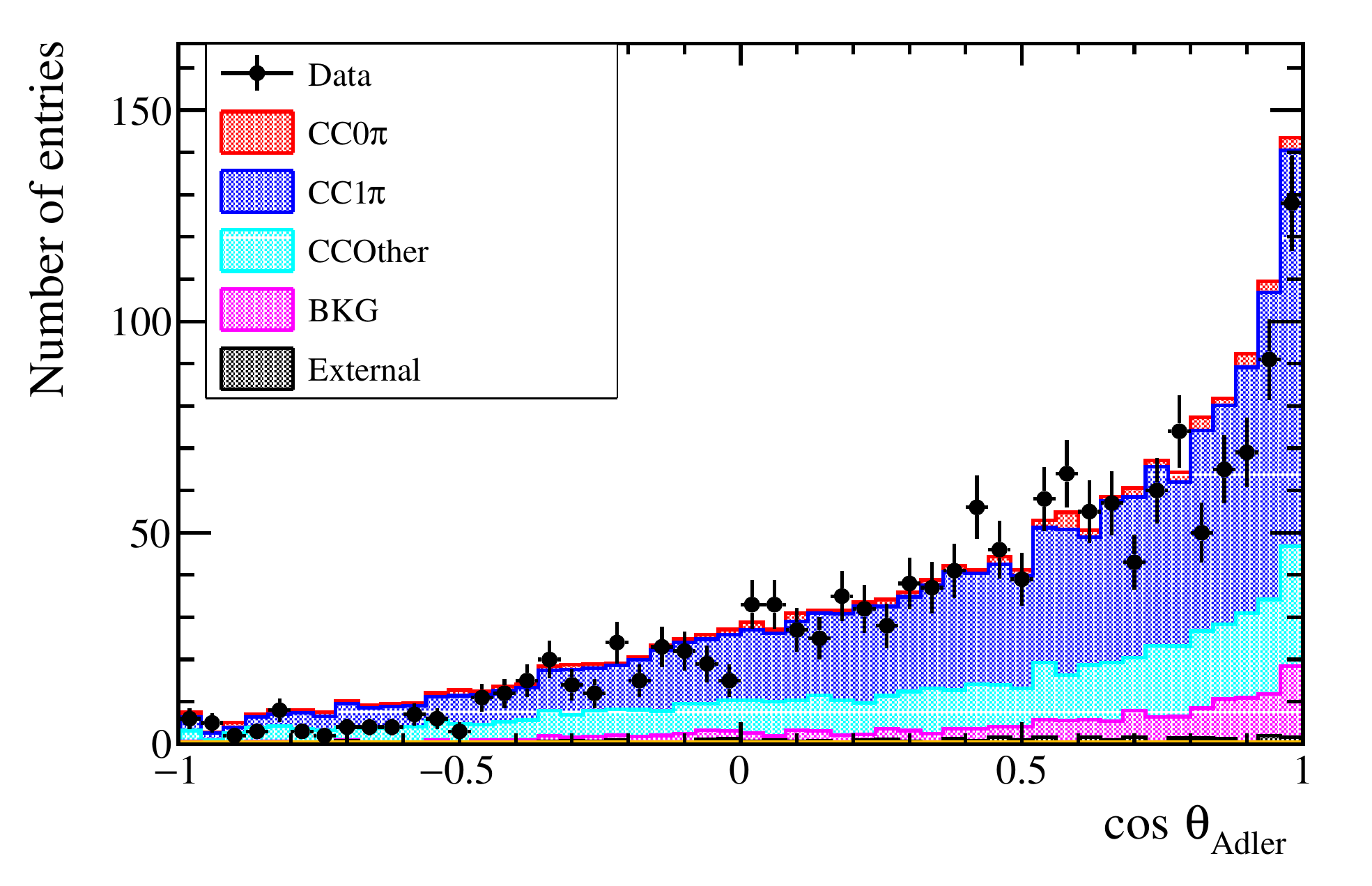}
  \includegraphics[width=.4\linewidth]{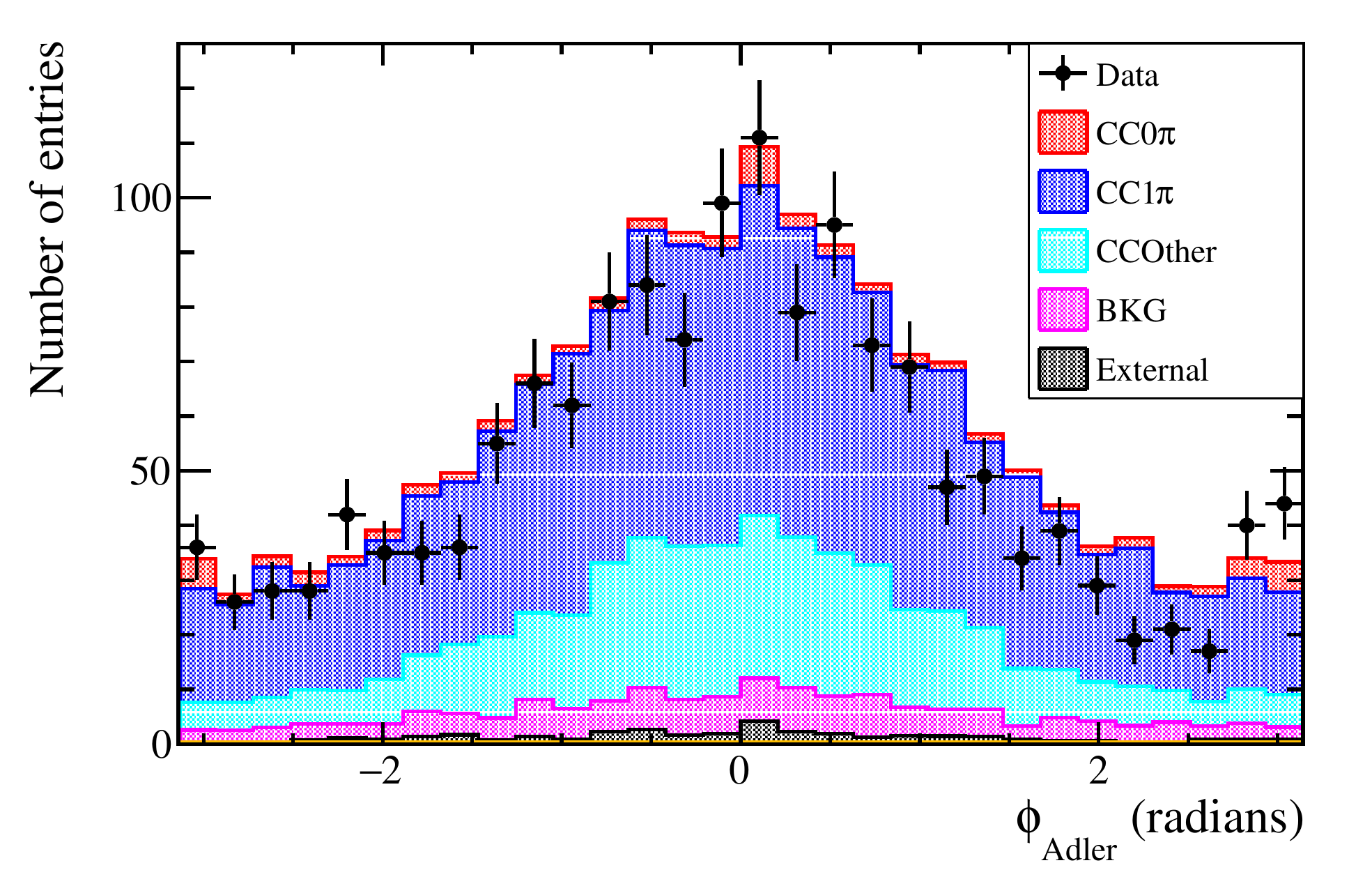}
  \caption{$\cos\theta_{Adler}$ and $\phi_{Adler}$ distributions for the selected \cconepi sample. Data are compared with NEUT 5.1.4.2.}
\label{fig:theta_Adler_Rec}
\end{figure*}

Existing models \cite{Hernandez:2007qq} predict an interference between the resonant and non-resonant pion production that leads to the transverse polarization as measured by the ANL data~\cite{ANL}. 
Figure~\ref{fig:theta_Adler_Rec} shows the distribution of  $\cos{\theta_{Adler}}$ (left) and $\phi_{Adler}$ (right)
for the sub-sample of \cconepi events with the pion reconstructed in the TPC.

\subsection{Control samples for background subtraction} \label{sec:controlsamples}

Control samples are selected in the data to constrain the normalization of several Monte Carlo background components listed in Table~\ref{tab:purity_CCsplited}.
Each control sample is selected to be representative of a specific background and it is required to minimize the content of \cconepi in order to be considered a 
side-band sample independent of the signal sample.
They are also required to be independent from each other. 
The three control samples, described in the following subsections, correspond one to CC$0\pi$ background and the other two to subsamples of the CCOther background, one with missing charged pions detection and the other with misidentified electrons or positrons.

For the contamination from interactions taking place out of the FGD fiducial volume, no control sample was found to reproduce the characteristics of this background. 
In this case the subtraction relies on the Monte Carlo prediction and the lack of a data constraint is taken into account in the systematic error estimation.
Control samples are used to extract normalization constants $\alpha_k={S_{\rm{data},k}}/{S_{\rm{MC},k}}$, where $S_{\rm{data},k}$ and $S_{\rm{MC},k}$ are the number of events in side-band $k$, respectively for data and Monte Carlo. 
These normalization constants $\alpha_k$ are used to re-scale the corresponding Monte Carlo background components before subtraction. The normalization constants are applied to each of the three background classes selected according to true Monte Carlo information.

\subsubsection {Control sample A}

\begin{figure*}[htb]
\centering
\includegraphics[width=.32\linewidth]{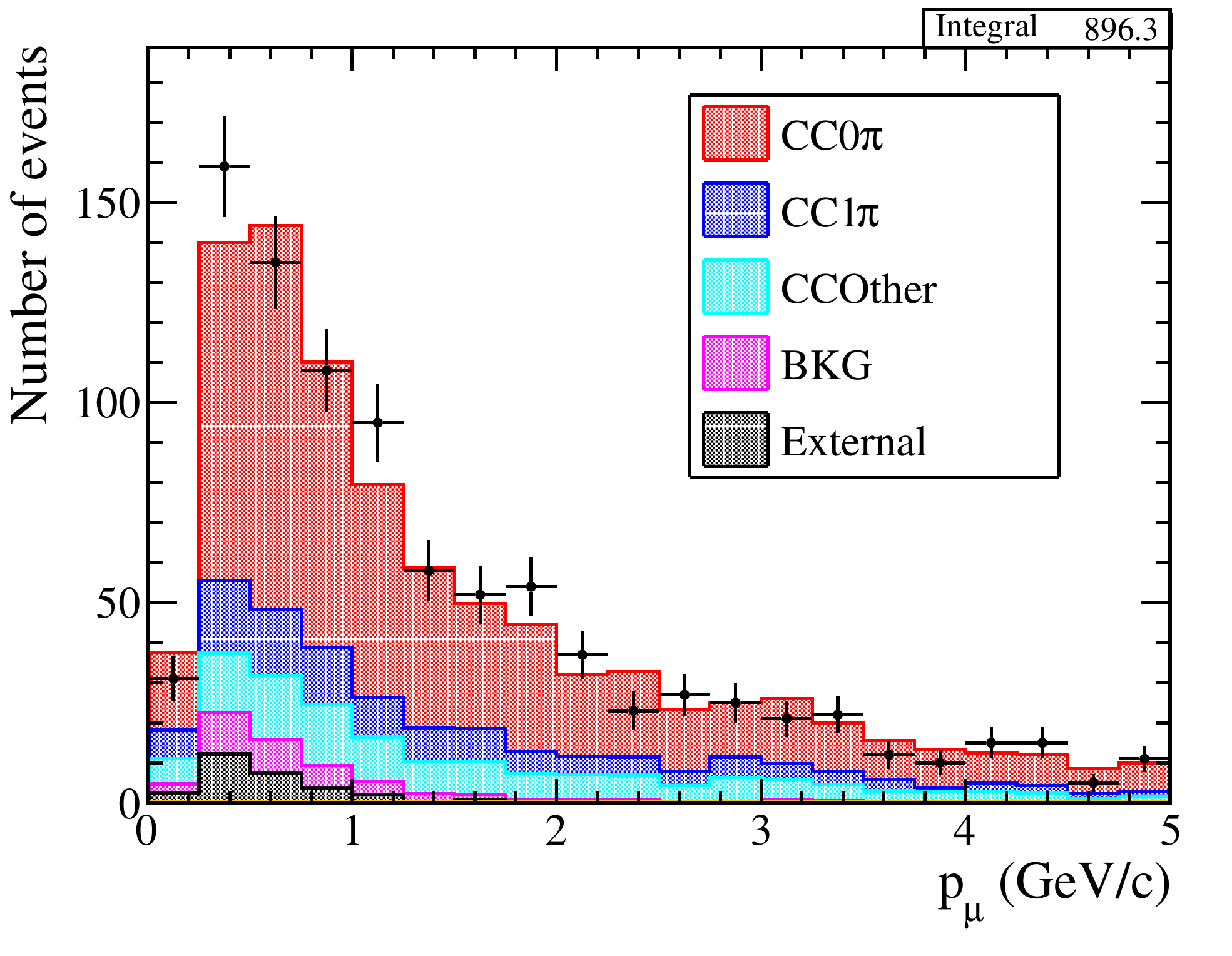}  
\includegraphics[width=.32\linewidth]{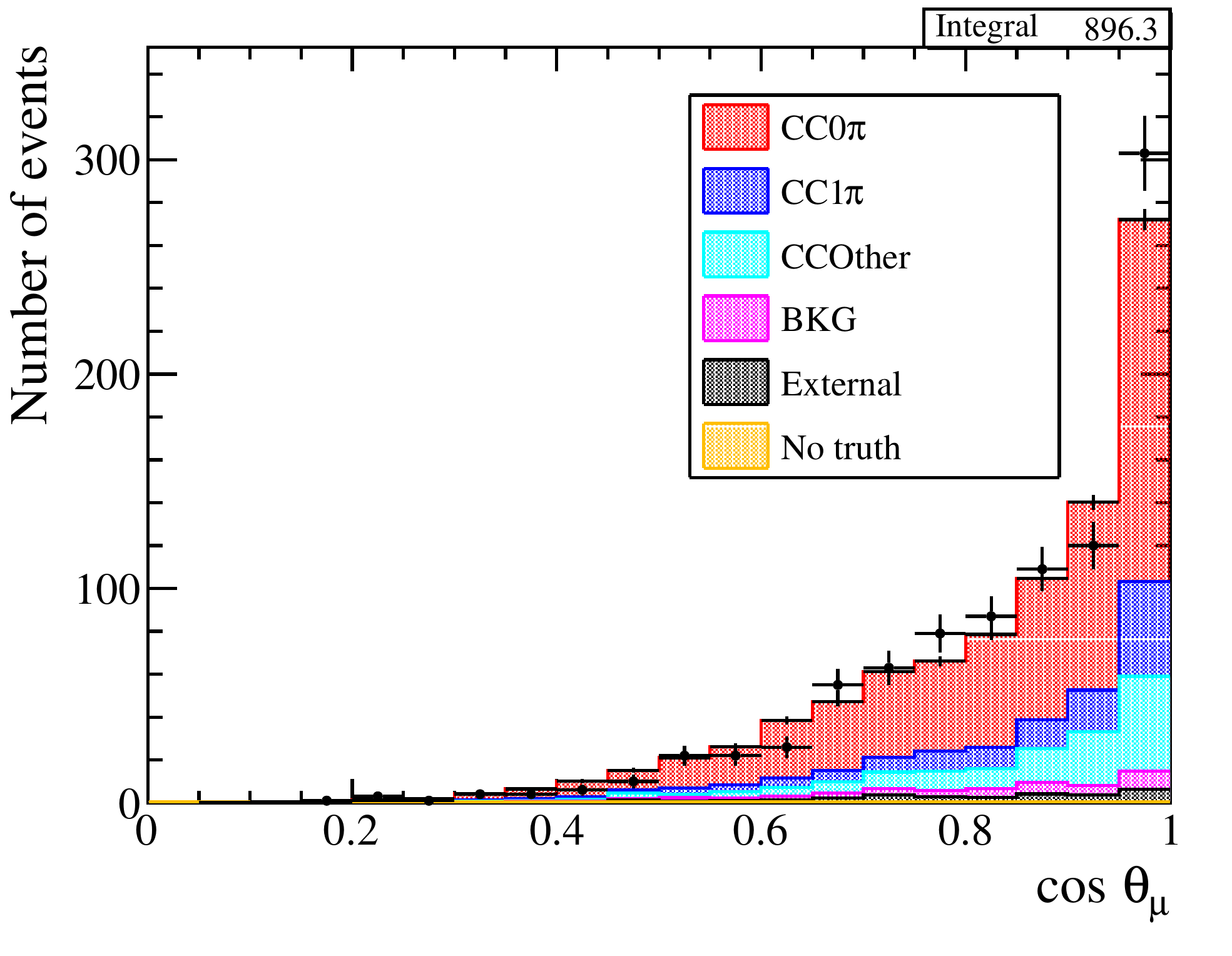} 
\includegraphics[width=.32\linewidth]{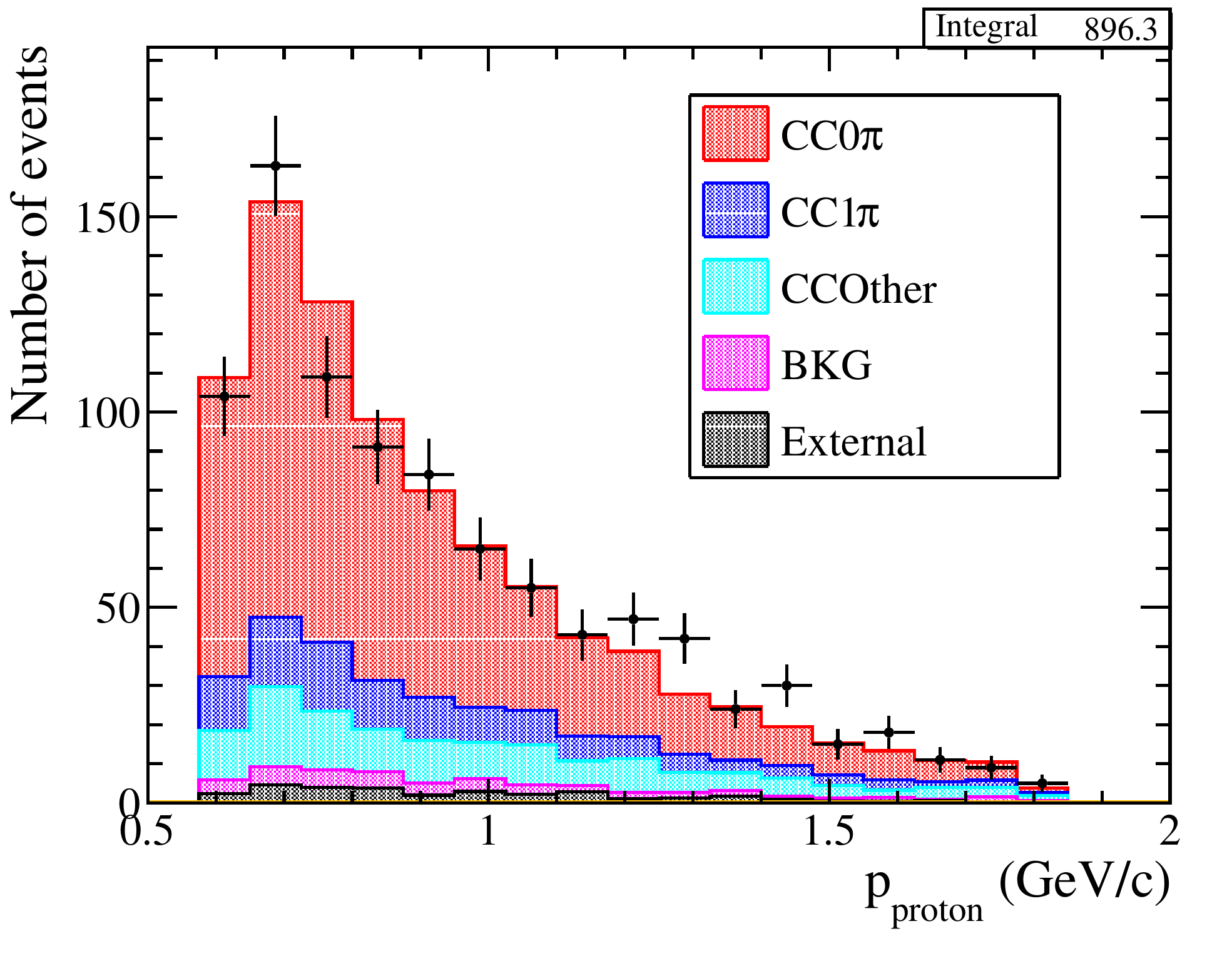} 
\caption{Control sample A: muon momentum (left) and cosine of the muon angle (middle) and proton momentum (right).}
\label{fig:CC0Pi1P}
\end{figure*}

One source of background is the CC$0\pi$ mis-identification, see Table~\ref{tab:purity_CCsplited}. Events where a proton is mis-identified as pion are a background in the \cconepi selection.  
The mis-identification arises from the similar ionization power of protons and pions around 1.5~GeV/c in the TPC.
The first control sample aims to select CC0$\pi$ events requiring a muon and no pions in the final state, with a proton identified in the final state and any number of additional nucleons~\cite{Abe:2015awa}. The selection requires one and only one additional TPC track, other than the muon track, with an energy deposition not compatible with a pion or an electron.
The angle with respect to the muon is required to be between 0.5 and 1.5~rad and the momentum must be between $0.6$ and $1.8$~GeV/c, which corresponds to the range where the mis-identification between pions and protons is larger. Table~\ref{tab:ControlSample_sideband} shows the topological composition of the control sample A. Figure~\ref{fig:CC0Pi1P} shows the muon candidate momentum and angle (top row) along with the proton momentum distribution for the selected sample in data and MC. From this control sample the extracted normalization value for the CC0$\pi$ contamination is $\alpha_A = 1.02$.

\begin{table}[htbp]
\caption{ Control samples composition. }
\label{tab:ControlSample_sideband}
\begin{center} 
\begin{tabular}{|l | c | c| c|}
\hline
   & \multicolumn{3}{c|}{Control samples} \\
   \cline{2-4}
   &  A &  B &  C \\
\hline
CC-0-pion & 64.2 \% & 1.9 \% & 4.6 \% \\
CC-1-pion & 13.7 \% & 22.0 \% & 8.4 \% \\
CCX$\pi^0$ & 9.5 \% & 40.0 \% & 56.2 \% \\
CCN$\pi^{+/-}$ & 5.1 \% & 21.7 \% & 12.2 \% \\
Non-$\nu_{\mu}CC$ & 4.1 \% & 10.7 \% & 6.9 \% \\
Out FGD1 FV & 3.4 \% & 3.8 \% & 11.8 \% \\
\hline
\end{tabular}
\end{center} 
\end{table}

\subsubsection {Control sample B}

\begin{figure*}[htb]
\centering
\includegraphics[width=.32\linewidth]{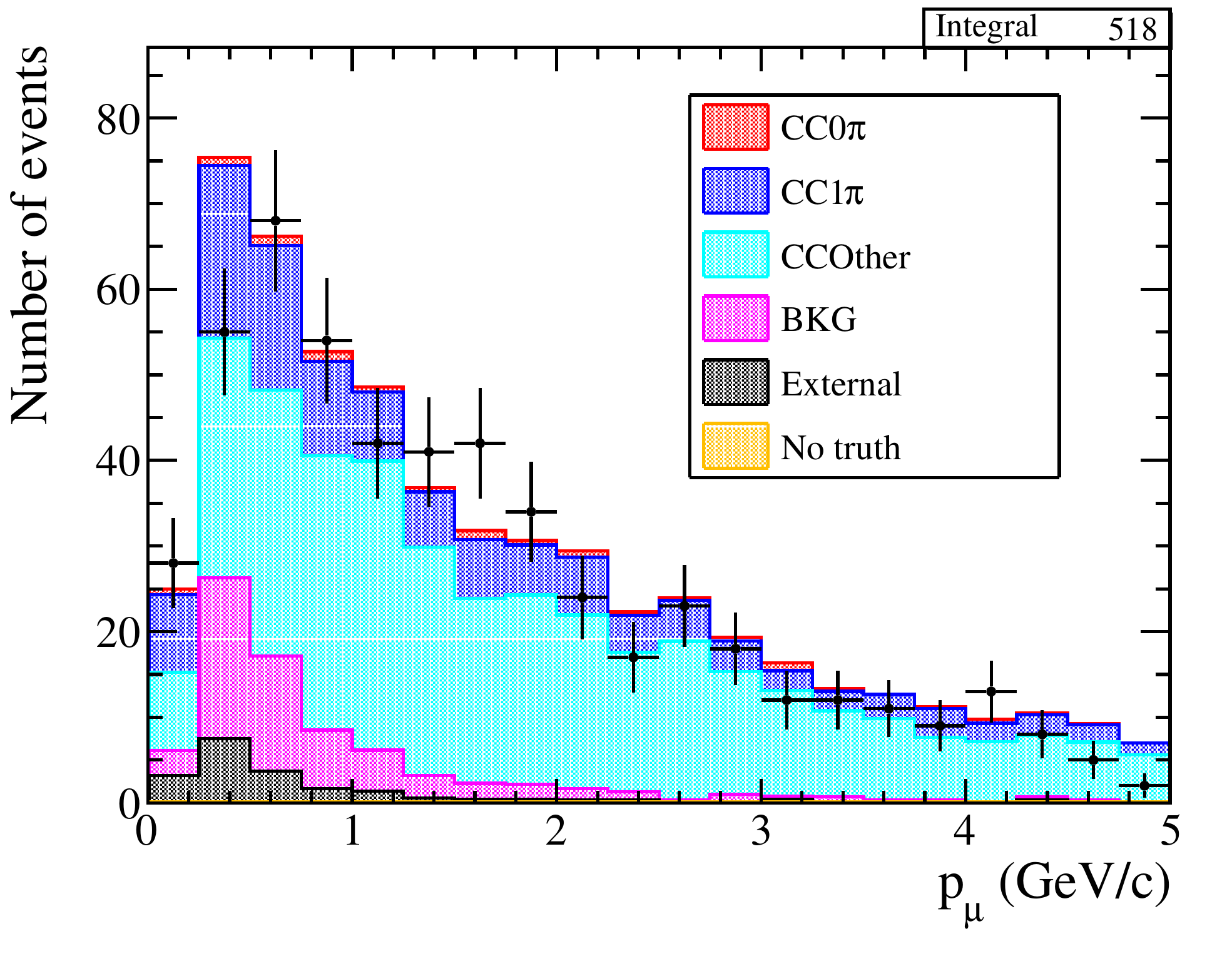}
\includegraphics[width=.32\linewidth]{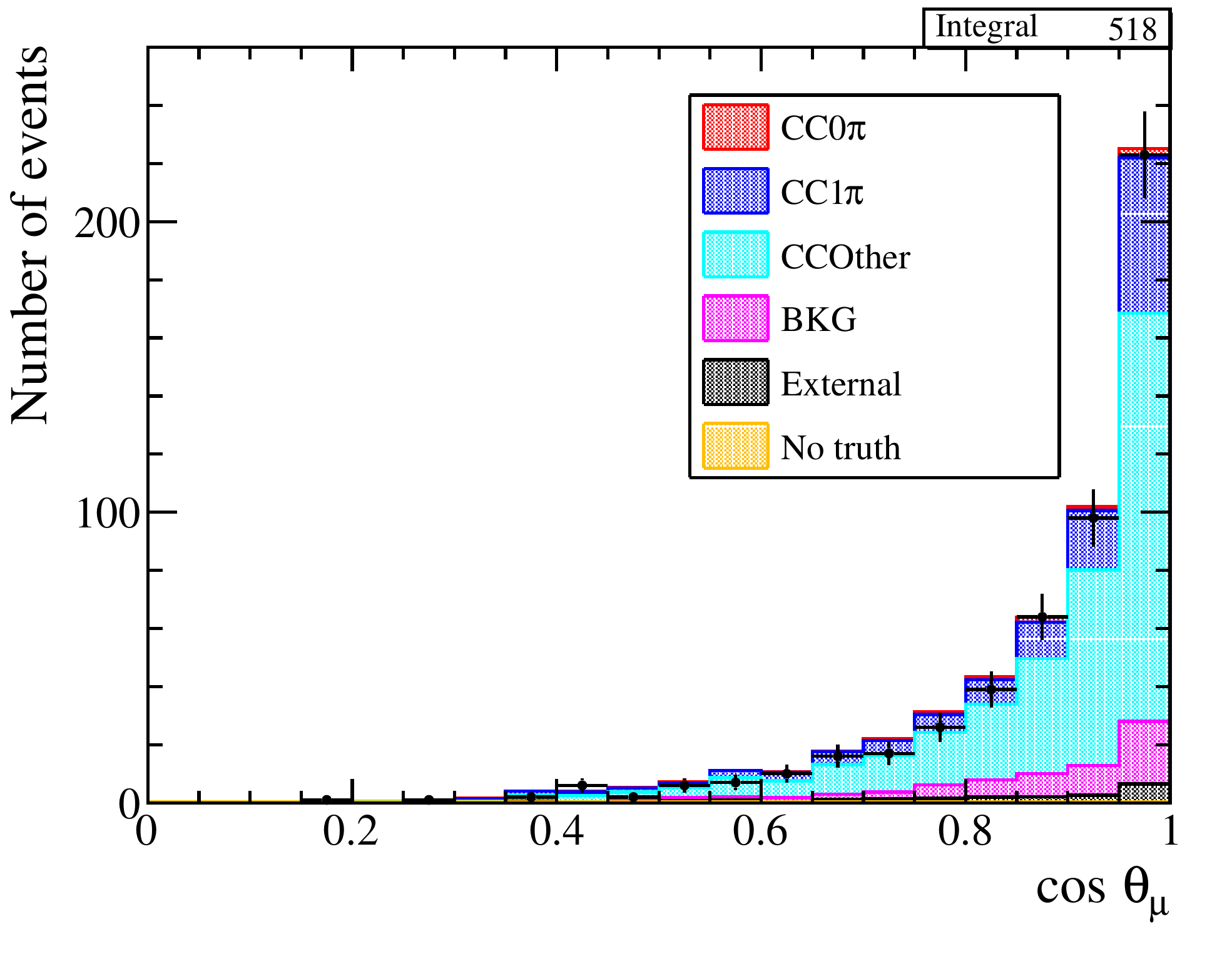}
\includegraphics[width=.32\linewidth]{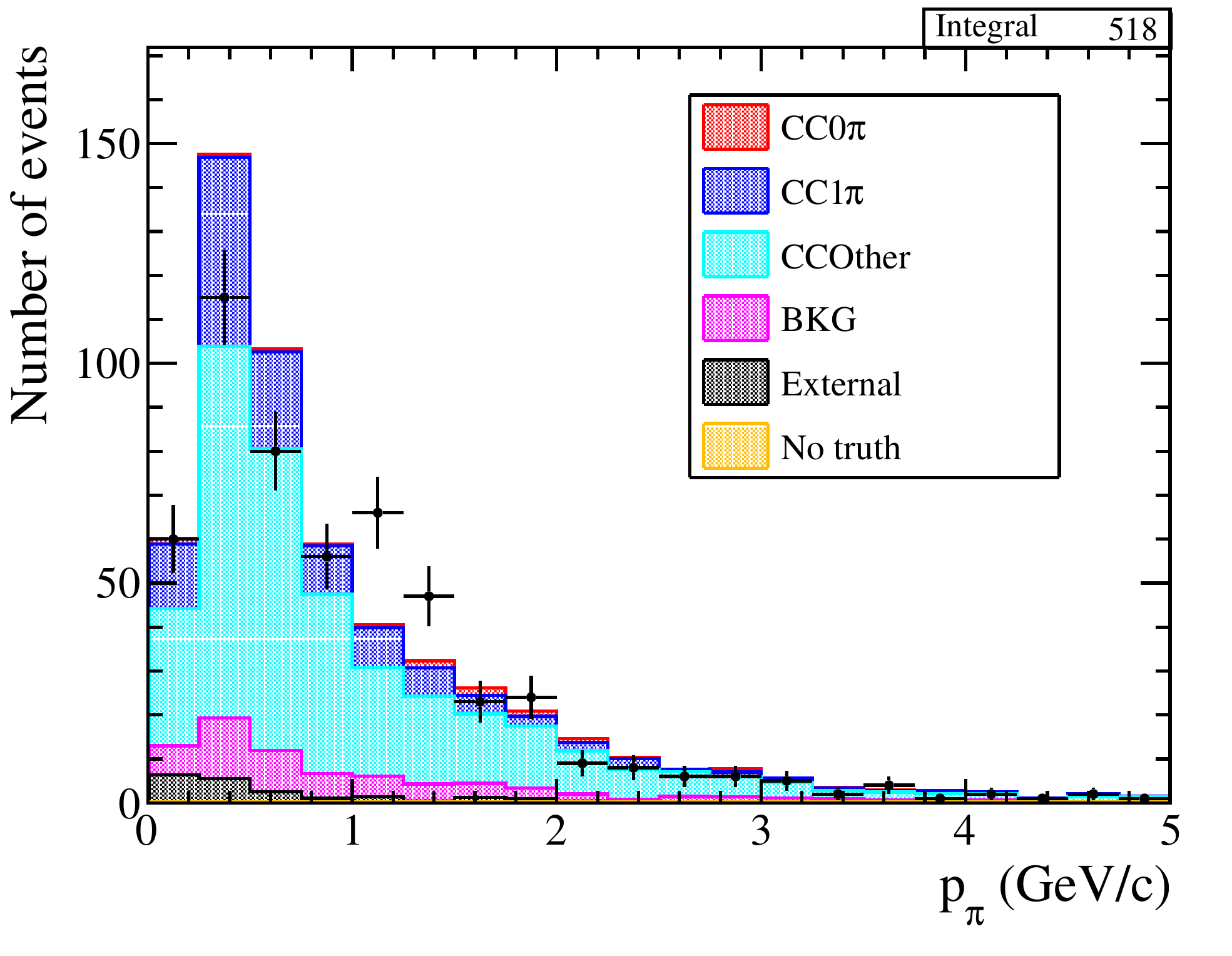}
\caption{Control sample B: Muon momentum (left) and angle (center) and pion momentum (right).}
\label{fig:controlB}
\end{figure*}

The second control sample is a subset of the CCOther sample, obtained requiring, besides the muon, the presence of two TPC tracks tagged as positively charged pions.
Events with three or more TPC tracks in addition to the muon track are rejected as they are high energy, multiple tracks events which are less representative of the actual backgrounds.
Table~\ref{tab:ControlSample_sideband} lists the topological composition of the control sample B. Figure~\ref{fig:controlB} shows the data and Monte Carlo comparison for the muon momentum and angle and for the pion momentum in this control sample.

This control sample is used to constrain the event contamination from CCOther due to multiple pion events. From this control sample we extract the normalization value $\alpha_B = 0.94$.

\subsubsection {Control sample C}

\begin{figure*}[htb]
\centering
\includegraphics[width=.4\linewidth]{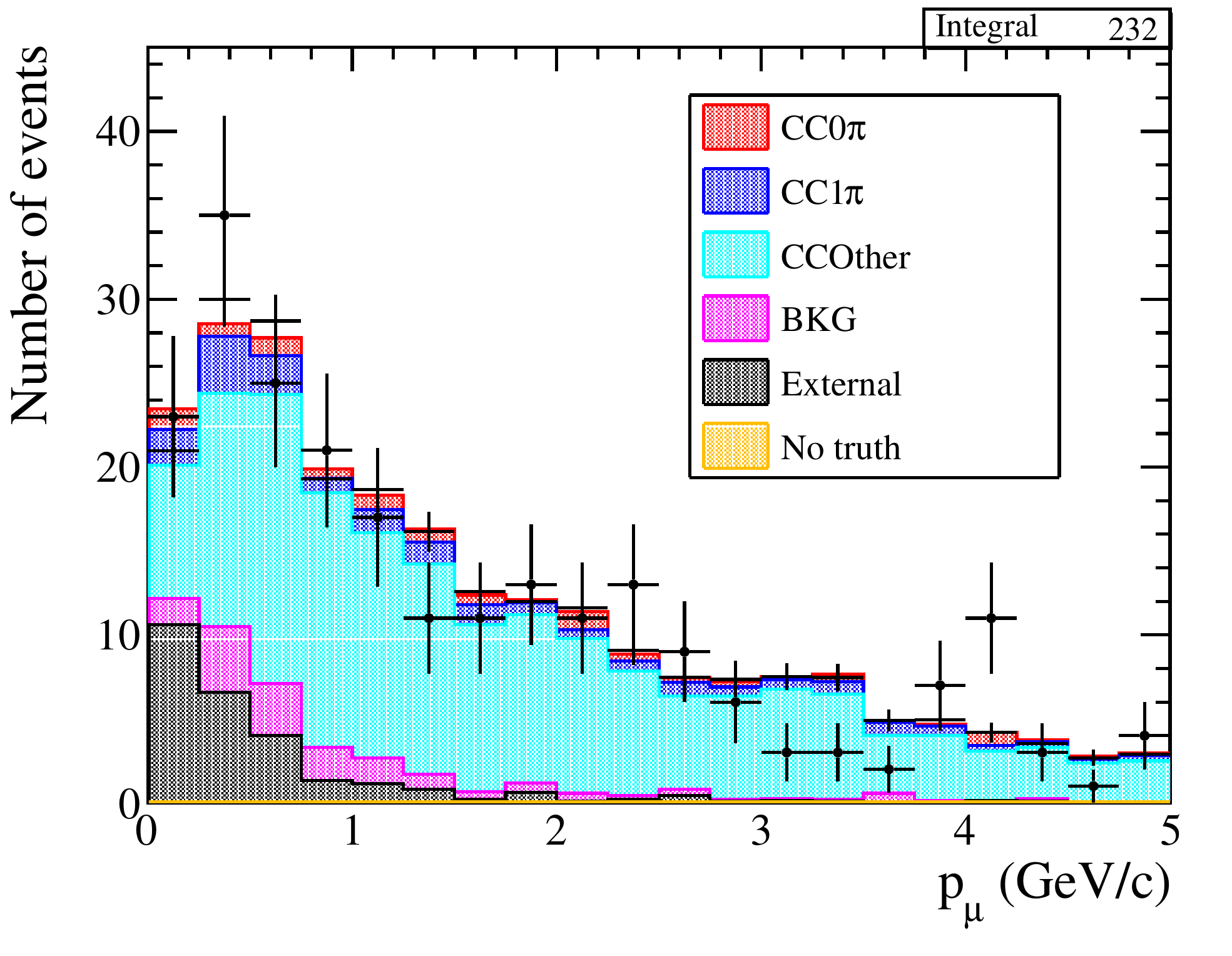}
\includegraphics[width=.4\linewidth]{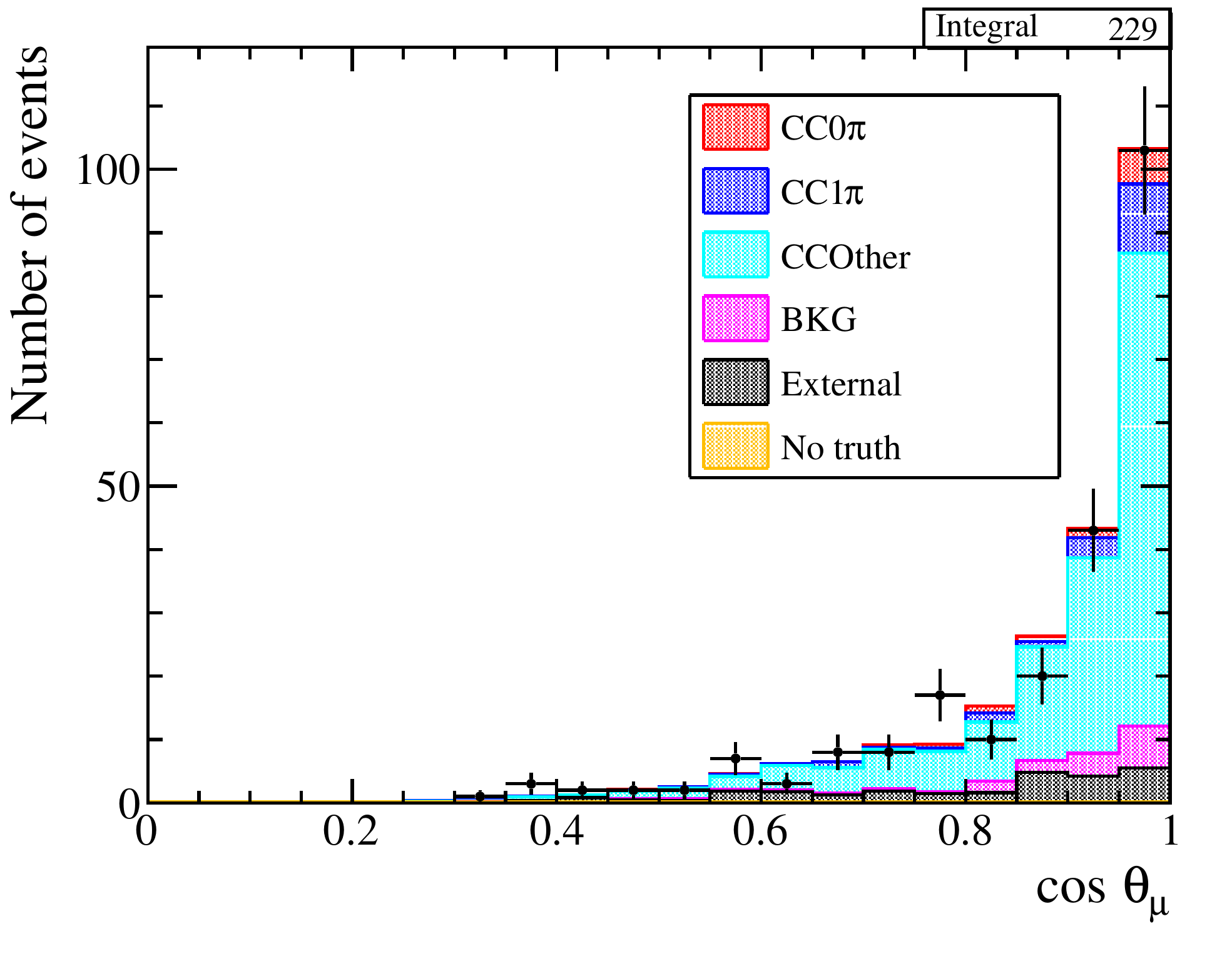}
\caption{Control Sample C: Muon momentum and angle distribution.}
\label{fig:controlC}
\end{figure*}

Similarly to the previous sample, this is a subset of the CCOther, obtained by additionally requiring at least one TPC track tagged as an electron or positron. This control sample is used to constrain the event contamination from CCOther due to neutral pion events.
To reject misidentified protons, positron tracks are required to have a momentum smaller than $0.4$~GeV/c. The absence of $\pi^+$ tracks in the TPC is required in order to avoid overlap with the control sample B.
The number of TPC tracks in addition to the muon track is required to be exactly two since with this requirement the shape of the control sample in Monte Carlo is found to be more similar to the actual background. Table~\ref{tab:ControlSample_sideband} provides the topological composition of control sample C.
Figure~\ref{fig:controlC} shows the comparison between data and Monte Carlo for the muon momentum and angle for this sample. The normalization value obtained for this control sample C is $\alpha_C = 0.99$.

\subsection{Systematic error}
\label{sec:detectorsystematics}
 
Systematic error can be split into three categories related to flux, detector and modeling of interactions. Detailed description of all the systematic errors can be found in Ref.~\cite{Abe:2015awa}. 

Uncertainty in the neutrino flux prediction arises from the hadron production model, proton beam profile, horn current, horn alignment, and other factors. For each source of uncertainty, the underlying parameters are varied to evaluate the effect on the flux prediction. The average effect of this systematic varies in the range 10-15\% along the different differential measurements.

The detector systematic errors are estimated by comparing the simulation predictions and dedicated data control samples.
This list of detector systematic errors include track efficiency in the FGD and TPC, particle identification, ECAL pion rejection, charge identification and momentum scale and resolution. The uncertainties caused by simultaneous events (pile-up), tracks coming from outside of the detector (sand muons), out-of-fiducial-volume (OOFV) events and secondary interactions (SI) of pions and nucleons in the detector are also evaluated. On average, the largest contribution from the detector systematics are the SI, while for low energies the charge mis-identification is dominant. 

A set of systematic parameters characterizes the uncertainties on the predictions of the NEUT generator. These uncertainties are propagated through the analyses to estimate the impact on the background and signal predictions, as well as the effect of the final state interactions. A number of those parameters are normalization uncertainties for the different interaction modes simulated by NEUT (energy dependent for the dominant modes at the T2K neutrino energy spectrum). Other parameters describe uncertainties on the values of the axial mass (using separate parameters for CCQE and resonant interactions), of the binding energy, and of the Fermi momentum. An additional systematic parameter covers the difference between the predictions obtained with the default relativistic Fermi gas model used by NEUT and a spectral function describing the momentum and energy of nucleons inside the nucleus \cite{Benhar:1995}. The modeling uncertainties are constrained by fits to external neutrino and pion scattering data (more details in Ref.~\cite{Abe:2015awa}).

Detector, beam and cross section model uncertainties were propagated in the selected sample.  The beam and the cross-section model uncertainties are propagated weighting the events according to true particle kinematics including the neutrino. Detector uncertainties are propagated event by event according to the observable on which the systematic depends; the algorithm depends on the systematic type. The propagation of the detector uncertainties is described in detail in~\cite{Abe:2015awa}.

All systematic uncertainties were propagated using a sample of toy experiments generated using the nominal values of each uncertainty. Each toy experiment is treated as data, i.e. the cross section is determined for each toy, and the results were used to calculate a covariance matrix defined as:
\begin{equation}
V_{ij} = \frac{1}{N} \sum_{s_n = 1}^{N} \left ( \sigma^{(s_n)}_i - \sigma^\text{nominal}_i \right ) \left ( \sigma^{(s_n)}_j - \sigma^\text{nominal}_j \right )
\label{eqn:analysis2:covarianceMatrix}
\end{equation}
where, for each source of uncertainty, labeled by $s$, two thousand pseudo-experiments are performed, giving a new differential cross section $\sigma^{(s_n)}$ each time, and the nominal cross section in bin $i$ is given by $\sigma^\text{nominal}_i$.

As an example of the effect of these systematics we show the impact of them in the double differential cross section measurement on the muon momentum and cosine of the angle, see Fig.\ref{fig:Covariance}. The systematic error contribution to this particular observable is 15.4\% from the beam flux uncertainty, 8.2\% from  the detector uncertainty and 8.7\% from the cross-section model uncertainties.

\subsection{Phase space}

The acceptance of the detector is limited in angle and momentum both for pions and muons. 
It is necessary to find suitable restrictions to identify the phase space where the observables can be unfolded without introducing large model dependencies. 
Complex kinematical observables (i.e. $Q^2$, $E_{\nu}$ and the Adler's angles) depend non-trivially on the ranges of angle and momentum of the selected particles.
We performed the phase space optimization independently for pions and muons.
The reconstruction efficiency has been studied both for the sub-sample of pions reconstructed in the TPC and the sub-sample of pions identified by the Michel electron tag. The resulting phase space 
for the reconstructed quantities is then associated with the true phase space contributing to the measured cross section.

The phase space for the muon observables is restricted to $\cos \theta_{\mu}> 0.2$ and $p_{\mu}> 0.2$~GeV/c. The same acceptance restrictions are applied 
to the pion observables: $\cos \theta_{\pi}> 0.2$ and $p_{\pi}>0.2$~GeV/c for charged pions with a TPC segment. In the cases when the pion is tagged with a Michel electron no pion phase space restriction is required. 
The bins in muon angle and momentum has been selected to ensure a large efficiency per bin while maintaining a large number of bins.
Table~\ref{table:Fiducial} summarizes the phase space restrictions applied for the differential cross section measurements presented in the next sections.

\begin{table*}
\caption{\label{table:Fiducial} Definition of the phase space restrictions used for the differential cross section measurements.}
\begin{ruledtabular}
\begin{tabular}{c|cccccc}
 Observable & $\cos{\theta_{\mu}} $  & $\cos{\theta_{\mu}}> 0.2 $  & $\cos{\theta_{\pi}}$ & $\cos{\theta_{\pi}}$ & $p_{\pi}$  &  Michel  \\
 & $ > 0 $ & $p_{\mu} > 0.2 $~$GeV/c$ & $> 0.2$ & $>0 $  & $>0.2$~$GeV/c$ &  Electron \\
\hline 
$d^2\sigma / d p_{\mu} d \cos{\theta_{\mu}} $ & Y & &  & & & Y \\
$d \sigma / d Q^ 2 $     & & Y & Y &   & Y  &  \\

$d\sigma / d p_{\pi} $   & & Y & Y &  & &  \\
$d\sigma / d \theta_{\pi} $ & & Y &    & Y  & Y &  \\
$d\sigma / d \theta_{\pi\mu} $ & &  Y &   Y &   & Y  &  \\
$d\sigma / d \phi_{Adler} $ & &  Y &  Y  &    & Y  &  \\
$d\sigma / d \theta_{Adler} $ & &  Y &  Y  &   & Y  &  \\
\end{tabular}
\end{ruledtabular}
\end{table*}

\section{Results}\label{sec:results}
 
The differential cross sections are extracted using the unfolding method proposed by D'Agostini~\cite{d'Agostini-unfolding}. The background prediction is subtracted from the data after they are weighted by the corresponding side-band normalisation ($\alpha_k$).  To assess the robustness of the method against potential biases, several tests were done using the nominal Monte Carlo to unfold pseudo experimental data produced with different Monte Carlo simulations obtained by changing the parameters and the models used to describe signal and background.
This study enables an understanding of the impact of the control samples and the optimal number of iterations needed for the unfolding procedure.
These tests show an optimal result with only one iteration. 
The binning for each observed variable was chosen taking into account the available statistics and the resolution of the reconstructed variables calculated using Monte Carlo simulation. 

\begin{figure*}
  \centering
  \includegraphics[width=0.99\linewidth]{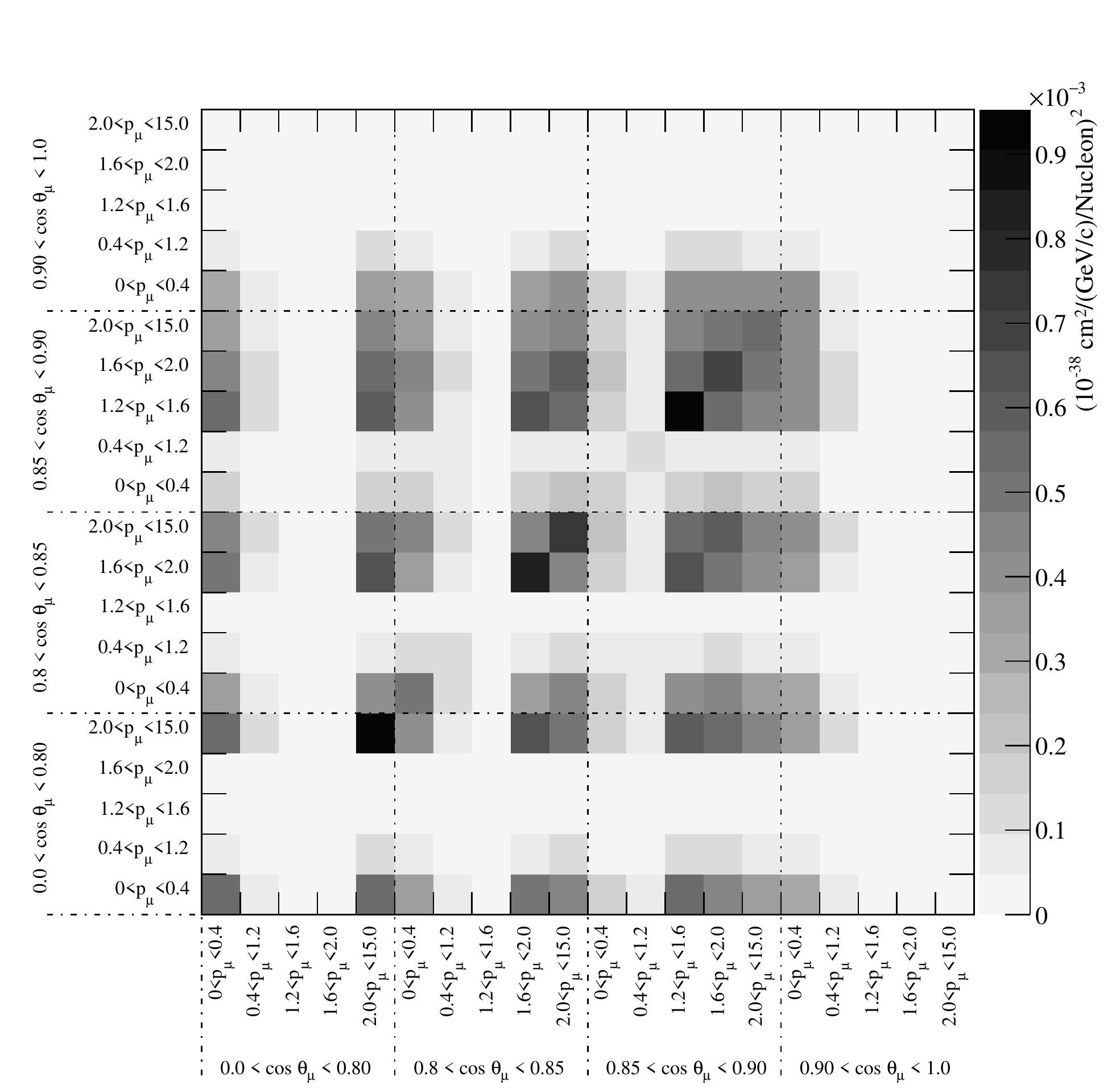}
  \caption{ Covariance matrix of the $d^2\sigma /
    dp_{\mu}d\cos\theta_{\mu}$ measurement in Figure~\ref{fig:PmuThetamu}. 
  }
  \label{fig:Covariance}
\end{figure*}

\begin{figure*}
  \centering
  \includegraphics[width=0.99\linewidth]{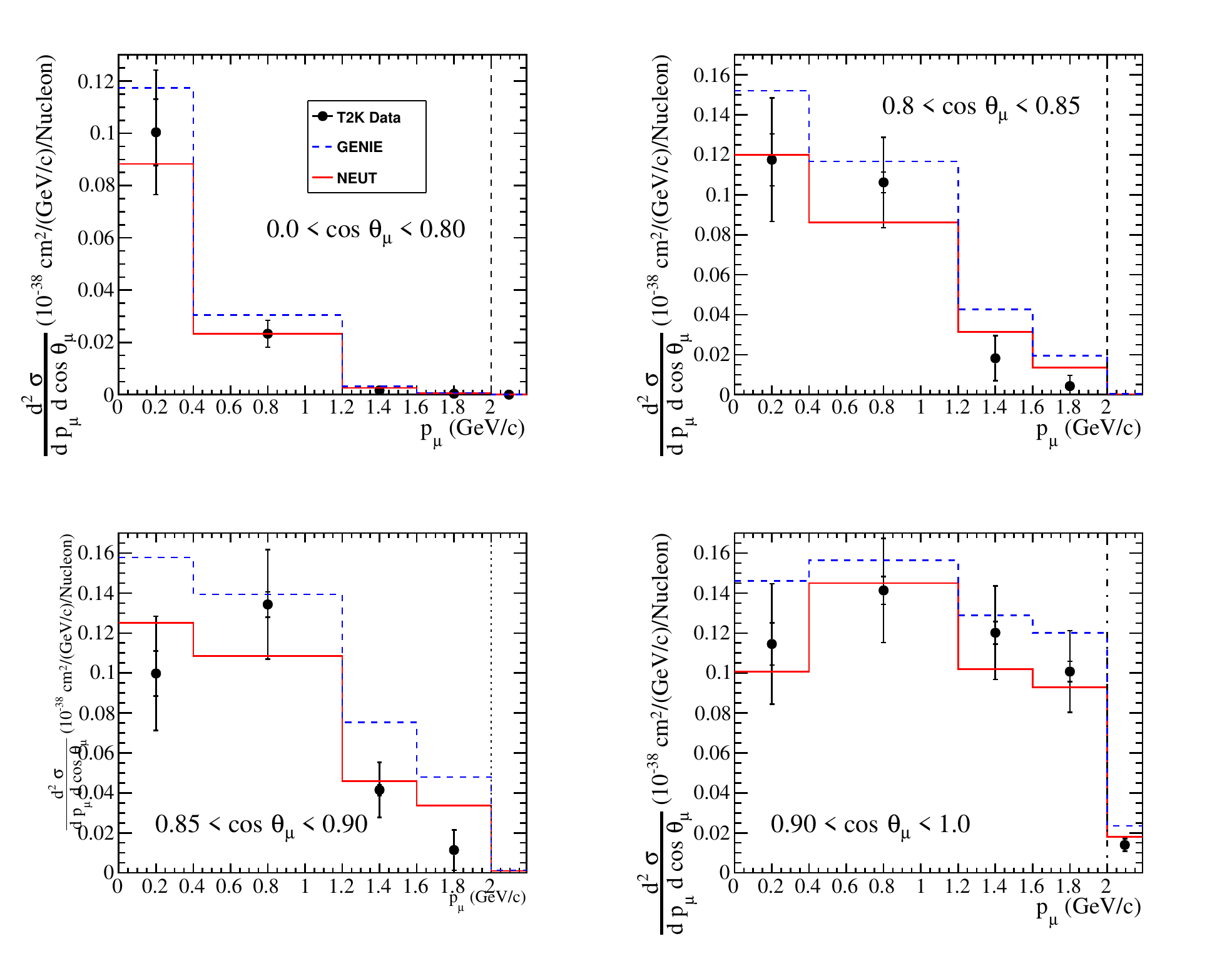}
  \caption{$d^2\sigma / dp_{\mu}d\cos\theta_{\mu}$ as a function of muon momentum for four cos$\theta_{\mu}$ 
  bins: (0.00,0.80) (upper left), (0.80,0.85) (upper right), 
  (0.85,0.90) (lower left), and (0.90,1.00) (lower right). The rightmost bin is truncated and it
   contains events up to 15~GeV/c in momentum. 
   The inner (outer) error bars show the statistical (total) errors. The lines show the NEUT (red) and GENIE (dashed blue) predictions.
  }
  \label{fig:PmuThetamu}
\end{figure*}

Figure~\ref{fig:PmuThetamu} shows the flux-integrated cross section for $d^2\sigma/dp_{\mu}d\cos\theta_{\mu}$. 
The rightmost bins are truncated and contain entries from 2 to 15~GeV/c. 
The unfolded double differential cross section as a function of the muon kinematics are well reproduced by the Monte Carlo given the large errors of this measurement, except some bin at high angles and momentum between 1.2 and 2.0~GeV/c. Figure~\ref{fig:Covariance} shows the covariance matrix including statistical and systematic errors.  

The result of the flux-averaged cross section value,
using the full sample, including the Michel electron tag, is:
$$\sigma = (11.76 \pm 0.44 \text{(stat)} \pm 2.39 \text{(syst)}) \times 10^{-40} \text{cm}^2\text{nucleon}^{-1}$$
while the corresponding value predicted by NEUT is $12.25 \times 10^{-40}{\rm cm}^2 /{\rm nucleon}$. The total cross section measurement depends strongly on model assumption for the extrapolation to the full phase space and it is provided only as a reference. 

\begin{figure}
  \centering
  \includegraphics[width=0.99\linewidth]{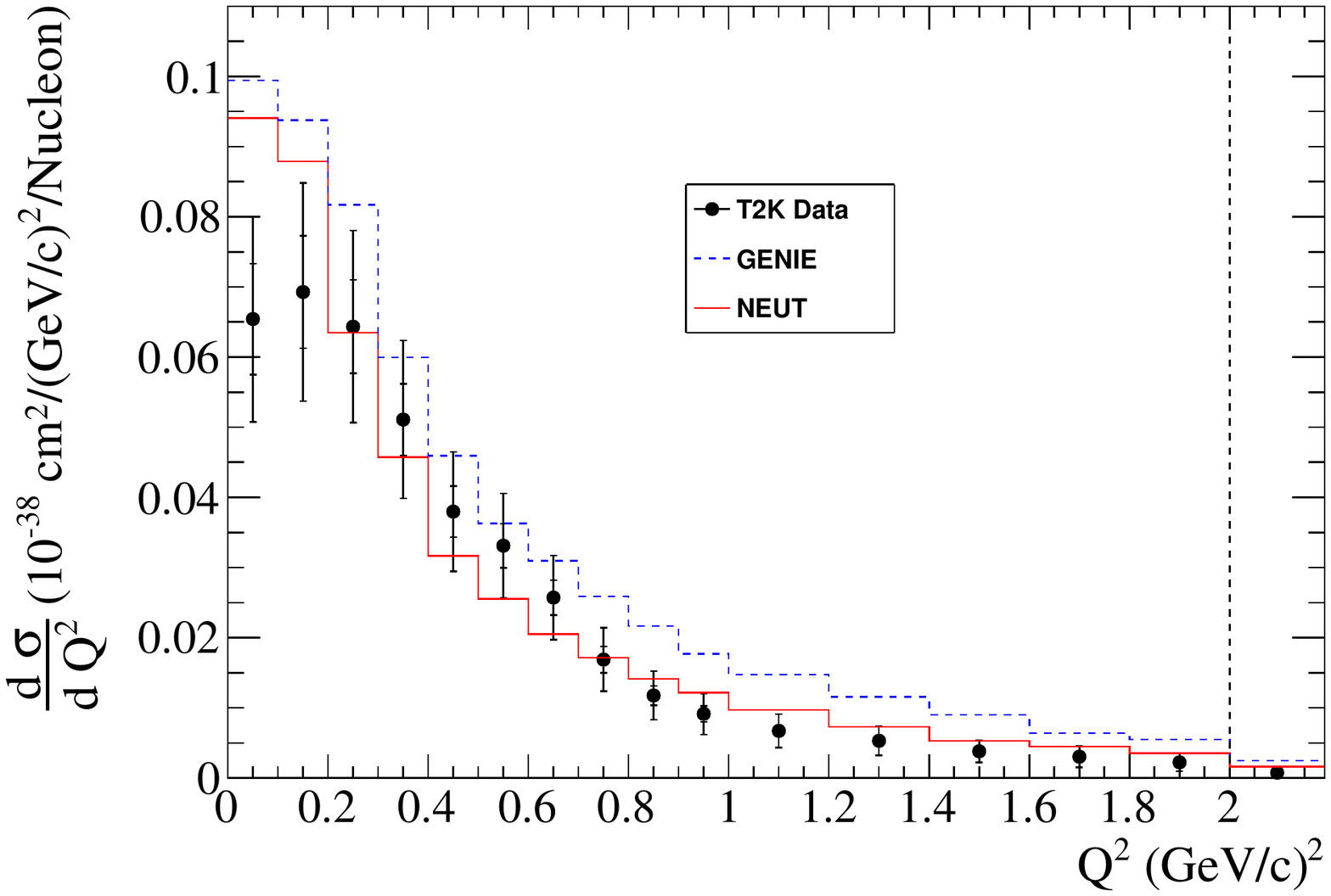}
\caption{$d\sigma / d Q^2$ differential cross section.
  The rightmost bin is truncated and it contains events up to 3.3~GeV$^2$/c$^2$.  
  The inner (outer) error bars show the statistical (total) errors. The lines show the NEUT (red) and GENIE (dashed blue) predictions. 
  }
  \label{fig:Q2}
\end{figure}

 Figure~\ref{fig:Q2} shows the unfolded $d\sigma/dQ^2$ flux-integrated cross section, measured in the restricted phase space of $\cos \theta_{\mu}>0.2$, $p_{\mu}>0.2$~GeV/c and $\cos \theta_{\pi}>0.2$, $p_{\pi}>0.2$ GeV/c. There is a significant difference in the shape of experimental results and the predictions. The pronounced model excess at low $Q^2$ might be an indication of deficiencies in the nuclear model.  

\begin{figure}
  \centering
  \includegraphics[width=0.99\linewidth]{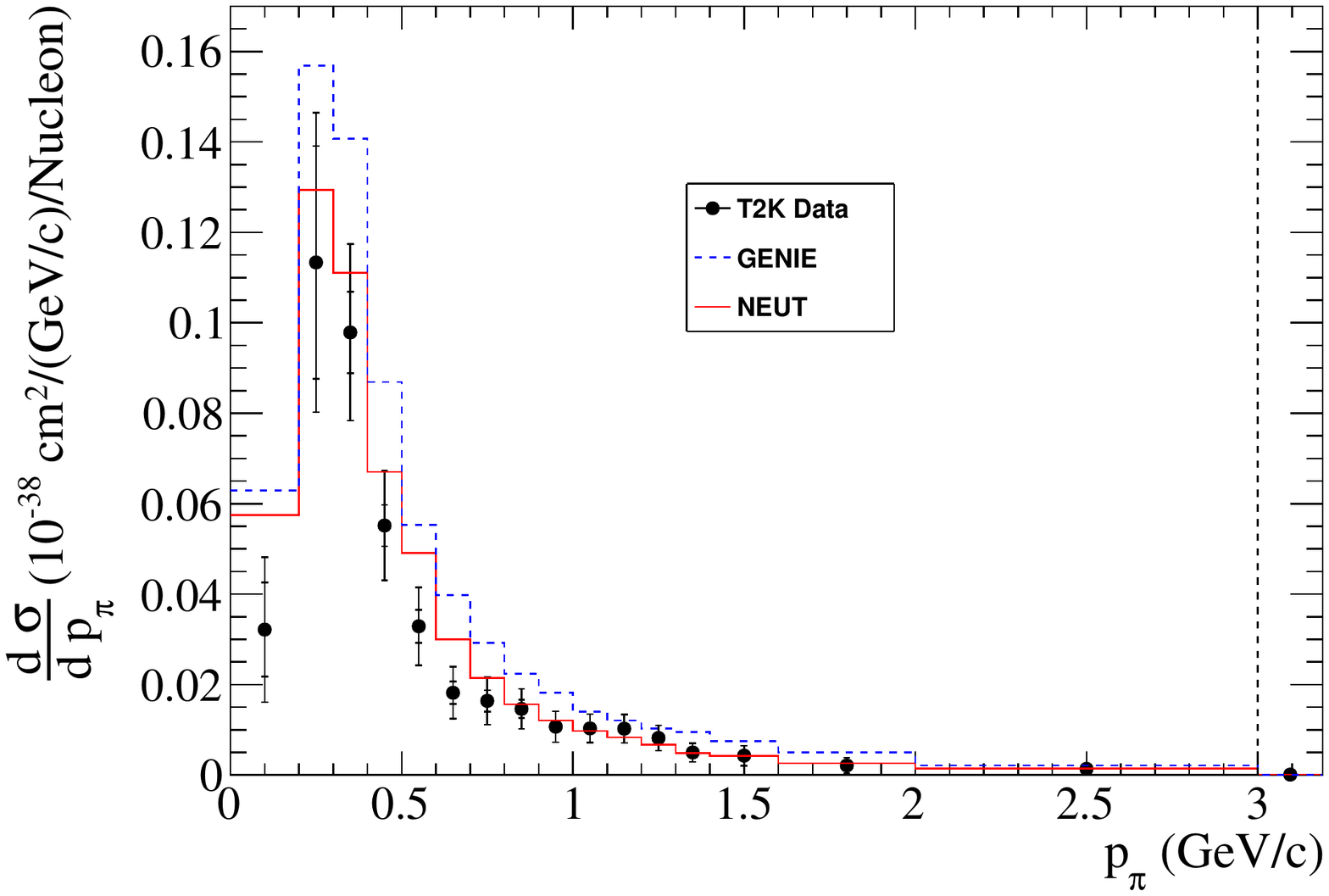}
  \caption{$d\sigma / d p_{\pi}$ differential cross section. 
    The rightmost bin is truncated and it contains events with momentum up to 15~GeV/c. 
    The inner (outer) error bars show the statistical (total) errors. The lines show the NEUT 5.1.4.2 (red) and GENIE 2.6.4 (dashed blue) predictions.
  }
  \label{fig:Ppi}
\end{figure}

Figure~\ref{fig:Ppi} shows the $d\sigma / d p_{\pi}$ flux-integrated cross section, measured in the restricted phase space of $\cos \theta_{\mu}>0.2$, $p_{\mu}>0.2$~GeV/c and $cos \theta_{\pi}>0.2$. 
Simulations overshoot data over the whole momentum range. NEUT shows a good agreement above 0.7~GeV/c. Similar model excess at low momentum pions has been observed in other experiments such as MiniBooNE~\cite{MiniBooNE-1pi}  and MINER$\nu$A~\cite{MINERvA-1pi,MINERvA-1pi_1,MINERvA-1pi_2}.

\begin{figure}
  \centering
 \includegraphics[width=0.99\linewidth]{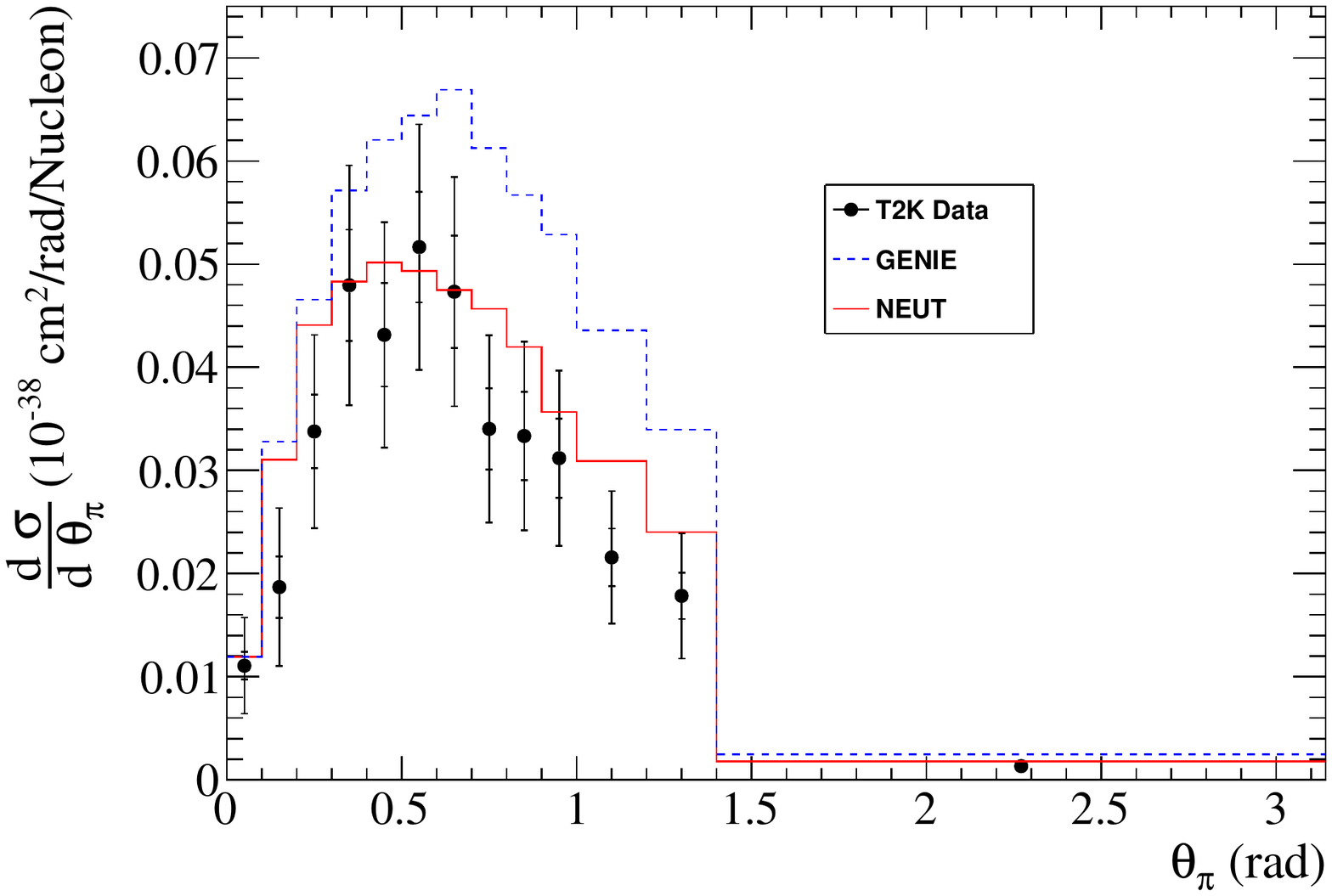}
  \caption{$d\sigma / d\theta_{\pi}$ differential cross section. 
    The inner (outer) error bars show the statistical (total) errors. The lines show the NEUT 5.1.4.2 (red) and GENIE 2.6.4 (dashed blue) predictions.
  }
  \label{fig:Thetapi}
\end{figure}

\begin{figure}
  \centering
 \includegraphics[width=0.99\linewidth]{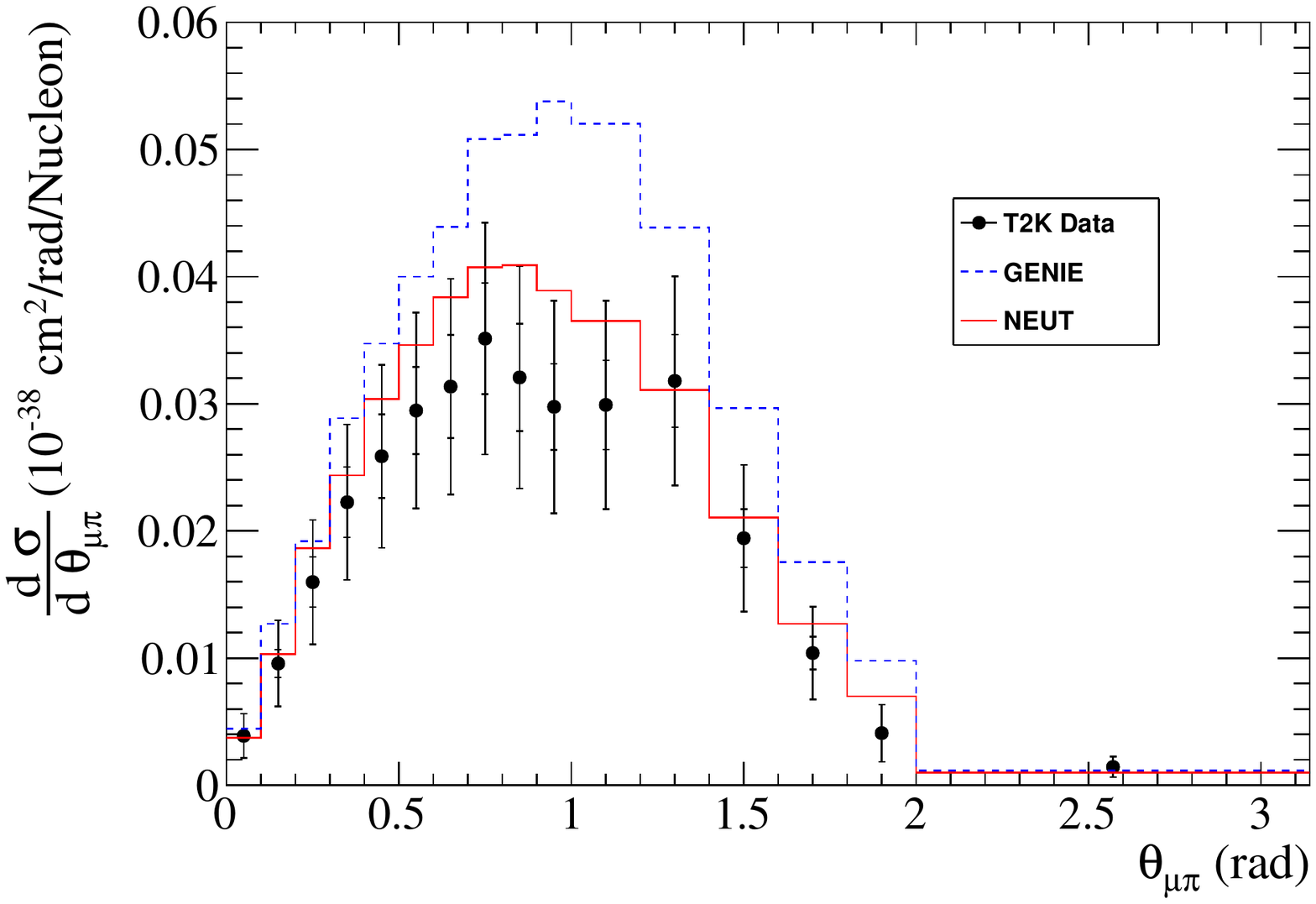}
\caption{ $d\sigma / d\cos\theta_{\mu\pi}$  differential cross section. 
    The inner (outer) error bars show the statistical (total) errors. The lines show the NEUT 5.1.4.2 (red) and GENIE 2.6.3 (dashed blue) predictions.
  }
  \label{fig:Thetapimu}
\end{figure}

 The $d\sigma/d\theta_{\pi}$ flux-integrated cross section is shown in Figure~\ref{fig:Thetapi}. The $\theta_{\pi}$ dependent cross section is measured in the restricted phase space $\cos \theta_{\mu}>0.2$, $p_{\mu}>0.2$~GeV/c for the muon and $\cos \theta_{\pi}>0$, $p_{\pi}>0.2$ for the pion. Consistently with the $d\sigma / dp_{\pi}$ cross section above, also the measured differential cross section as a function of the pion angle shows a disagreement with the predictions. 
 Figure~\ref{fig:Thetapimu} shows the $d\sigma/d\theta_{\pi \mu}$ flux-integrated
cross section, measured in the restricted phase space $\cos \theta_{\mu}>0.2$,
$p_{\mu}>0.2$~GeV/c for the muon and $\cos\theta_{\pi}>0.2$, $p_{\pi}>0.2$ GeV/c for
the pion. 

\begin{figure}
  \centering
    \includegraphics[width=0.99\linewidth]{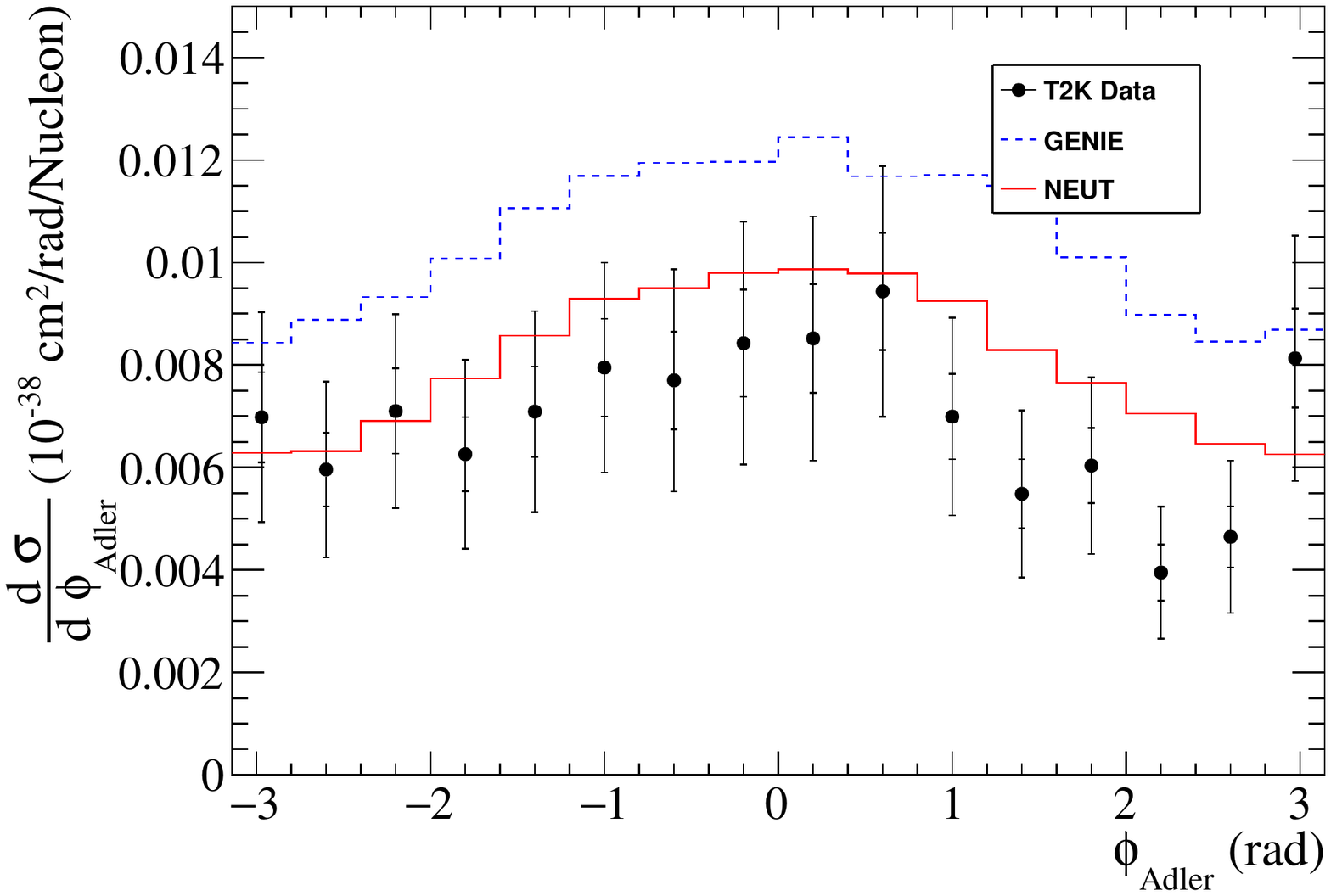}
  \caption{$d\sigma / d \phi_{Adler}$ differential cross section.
    The inner (outer) error bars show the statistical (total) errors. The lines show the NEUT 5.1.4.2 (red) and GENIE 2.6.3 (dashed blue) predictions.
}
  \label{fig:Phiplanar}
\end{figure}

Figure~\ref{fig:Phiplanar} shows the $d\sigma/d\phi_{Adler}$ flux-integrated cross section, measured in the restricted phase space of $\cos \theta_{\mu}>0.2$, $p_{\mu}>0.2$~GeV/c and $\cos \theta_{\pi}>0.2$, $p_{\pi}>0.2$ GeV/c.  The shape of the distribution is reasonably described by NEUT except for those values in between 0.8 and 2.8 rad. 
The region with the largest data deficit is around $\phi_{Adler} \simeq 1.5$ similar to the deficit observed in ANL data around $\pi/2$ for charged pions~\cite{ANL} and around the same value for neutral pions in MINER$\nu$Aa\cite{Minervapi0}.
A significant difference of the ANL measurement compared to T2K is the use of a deuterium target where both the Fermi momentum and the FSI are reduced with respect to the CH target. The ratio of the integrated cross-section for positive $\phi_{Adler}$ angles over the negative $\phi_{Adler}$ angles, similar to the Left-Right measured in MINER$\nu$A~\cite{Minervapi0}, gives a value of: $1.08\pm0.10$. NEUT and GENIE generators predict a value equal to 1. Both generators predictions show an unexpected dependency with the $\phi_{Adler}$ angle, see Fig.~\ref{fig:Phiplanar}. most probably caused by the effect of intra-nuclear cascade (FSI) on the reconstruction of the Adler's reference system~\cite{Sanchez:2015yvw}.

\begin{figure}
  \centering
    \includegraphics[width=0.99\linewidth]{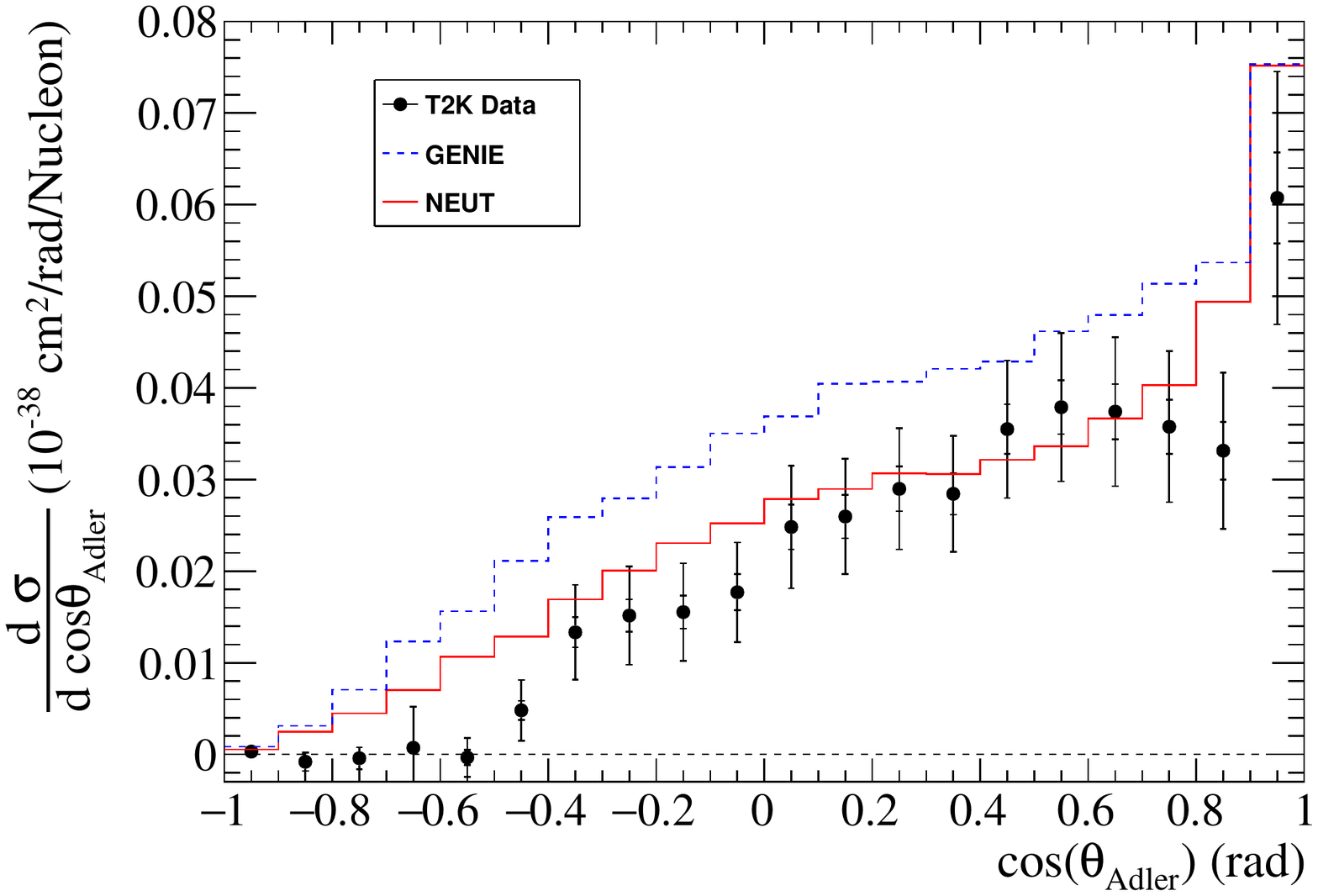}
  \caption{$d\sigma / d\cos\theta_{Adler}$ differential cross section.
    The inner (outer) error bars show the statistical (total) errors. The lines show the NEUT 5.1.4.2 (red) and GENIE 2.6.4 (dashed blue) predictions.
  }
  \label{fig:Thetaplanar}
\end{figure}

The experimental results are consistently below the NEUT prediction for negative values of $\cos\theta_{Adler}$, see Fig.~\ref{fig:Thetaplanar}. 
Negative $\cos\theta_{Adler}$ corresponds to low momentum pions ($\le 0.3$~GeV/c). 
This observation is consistent with the prediction excess observed at low pion momentum, see Fig.\ref{fig:Ppi}.

While the Monte Carlo reproduces reasonably well the muon observables, the predictions for the pion observables are larger than data. The difference between the two is the inclusion of Michel electron tags for the muon only observables. The difference might be an indication of a biased estimation of the Michel electron tagging efficiency in Monte Carlo, but also Final State Interaction modelling or the model prediction for the pion momentum will contribute to the observed disagreement. Even if the number of events were similar, there are significant shape differences in most of the obervables investigated. 

\section{Conclusions\label{sec:conc}}

The analysis presented in this paper describes the CC$1\pi^+$ cross section measurement on CH realized in the ND280, the off-axis near detector of the T2K experiment.

Using NEUT as the default MC generator we observe a purity of the CC$1\pi^+$ signal of 
$61.5\%$.
The main contamination in the sample is due to unidentified CCOther events. Three control
samples have been investigated in order to subtract the background using data instead of applying the Monte Carlo purity correction, with the aim to reduce the model dependency.

The aim of the distributions presented here is to provide results as much as possible in a model independent way, to make their comparison 
to other experiments easier and to contribute to the improvement of current models. 

We have presented differential cross section measurements using a set of observables that will be most useful for comparison with neutrino interaction models. 
One example is the use of Adler angles,
which have not been measured since the bubble chamber experiments \cite{ANL,BNL}.  
This is the first time those angles are measured in
interactions of neutrinos on heavy nuclei. 

The largest contribution to the measurement error overall is the uncertainty on the flux, while the largest contribution from detector systematics comes from pion secondary interactions and, at low energies, from the TPC charge mis-identification. Uncertainties in the cross section model 
are the second largest contribution to the uncertainties, which serves as a reminder of the importance of cross section measurements.

From the differential cross section measurements presented we highlight the following:

\begin{enumerate}
\item We observe a good description of the data for the
  CC$1\pi^+$ topological channel in all the muon kinematics observables.
These distributions use inclusively all pions, including the low energy pions identified by Michel electron tagging.
\item The shape of the predicted $Q^2$ distribution shows large discrepancies with data all over the available $Q^2$  space being more pronounced for $Q^2 \le 0.3 $~GeV$^2$/c$^2$.

\item We observe, in general, that the models predict larger cross sections for the angular pion observables. Only pions with momentum above 0.2~GeV/c, which have been identified as tracks in the TPC, are included. 
  The discrepancy is more pronounced for low momentum pions and are almost independent of the value of the $\theta_{\pi}$ and $\theta_{\mu\pi}$  angles. 
  
\item The MC model appears to predict larger number of events tagged by a Michel electron and smaller number of events with pions above 0.2~GeV/c (TPC tagged) than the rates observed in the experiment. The sum of both the TPC and the Michel electron samples show a reasonable agreement with both generator predictions. The observed disagreement might be caused either by a distorted pion momentum spectrum or by deficiencies in the efficiency predictions.  

\end{enumerate}

 We have also computed the flux-averaged  cross section value:

$$\sigma = (11.76 \pm 0.44 \text{(stat)} \pm 2.39 \text{(syst)})\times 10^{-40} \text{cm}^2\text{nucleon}^{-1}$$

To obtain this value 
the full \cconepi candidate sample is considered, including pions identified by the Michel electron tag. From this result
we extrapolated to the full phase space, including regions where the detector efficiency is small or even null, 
this result is strongly dependent on model assumptions and should be used with care.

\begin{acknowledgments} 
We thank the J-PARC staff for superb accelerator performance. We thank the CERN NA61/SHINE Collaboration for providing valuable particle production data. We acknowledge the support of MEXT, Japan; NSERC (Grant No. SAPPJ-2014-00031), NRC and CFI, Canada; CEA and CNRS/IN2P3, France; DFG, Germany; INFN, Italy; National Science Centre (NCN) and Ministry of Science and Higher Education, Poland; RSF, RFBR, and MES, Russia; MINECO and ERDF funds, Spain; SNSF and SERI, Switzerland; STFC, UK; and DOE, USA. We also thank CERN for the UA1/NOMAD magnet, DESY for the HERA-B magnet mover system, NII for SINET4, the WestGrid and SciNet consortia in Compute Canada, and GridPP in the United Kingdom. In addition, participation of individual researchers and institutions has been further supported by funds from ERC (FP7), ``la Caixa” Foundation (ID 100010434, fellowship code LCF/BQ/IN17/11620050), the European Union’s Horizon 2020 Research and Innovation programme under the Marie Sklodowska-Curie grant agreement no. 713673 and H2020 Grant No. RISE-GA644294-JENNIFER 2020; JSPS, Japan; Royal Society, UK; the Alfred P. Sloan Foundation and the DOE Early Career program, USA.
\end{acknowledgments}


\begin{thebibliography}{9}

\bibitem{ANL}  G. M.~Radecky, V. E.~Barnes, D. D.~Carmony, A. F.~Garfinkel, M.~Derrick, E.~Fernandez, L.~Hyman, G.~Levman, D.~Koetke  {\it et al.}, {\it Study of single-pion production by weak charged currents in low-energy $\nu -d$ interactions}, Phys. Rev. D {\bf 25}, 5 (1982) 1161.

\bibitem{BNL} T.~Kitagaki, H.~Yuta, S.~Tanaka, A.~Yamaguchi, K.~Abe, K.~Hasegawa, K.~Tamai, S.~Kunori,  Y.~Otani {\it et al.}, {\it Charged Current exclusive pion production
in neutrino deuterium interactions}, Phys. Rev. D {\bf 34}, 5 (1986) 2554.

\bibitem{K2K-cc1pr} A.~Rodriguez {\it et al.}(K2K Collaboration), {\it Measurement of single charged pion production in the charged-current interactions of neutrino in a 1.3~GeV wide band beam}, Phys. Rev. D {\bf 78}, (2008) 032003.

\bibitem{K2K-nc0pr} S.~Nakayama {\it et al.} (K2K Collaboration), {\it Measurement of single $\pi^\circ$ production in neutral current neutrino interactions with water by a 1.3~GeV wide band muon neutrino beam}, Phys. Lett. B {\bf619} (2005) 255

\bibitem{MiniBooNE-ratio} A.A.~Aguilar-Arevalo {\it et al.} (MiniBooNE Collaboration), {\it Measurement of the ratio of the ${\ensuremath{\nu}}_{\ensuremath{\mu}}$ charged-current single-pion production to quasielastic scattering with a 0.8 GeV neutrino beam on mineral oil}, Phys. Rev. Lett. 103 (2009) 081801.

\bibitem{ANL-ratio} M.~Derrick, E.~Fernandez, L.~Hyman, G.~Levman, D.~Koetke, B.~Musgrave, P.~Schreiner, R.~Singer, A.~Snyder  {\it et al.}, {\it Study of single-pion production by weak neutral currents in low-energy $\ensuremath{\nu}d$ interactions}, Phys. Rev. D {\bf 23} (1981) 569.

\bibitem{MiniBooNE-1pi} A.A.~Aguilar-Arevalo {\it et al.} (MiniBooNE Collaboration), {\it Measurement of Neutrino-Induced Charged-Current Charged Pion Production Cross Sections on Mineral Oil at $E_{\nu}\sim 1$ GeV}, Phys. Rev. D {\bf 83}, 052007 (2011).

\bibitem{MINERvA-1pi} B.~Eberly {\it et al.} (MINER$\nu$A Collaboration), {\it Charged Pion Production in $\nu_{\mu}$ Interactions on Hydrocarbon at $<E_{\nu}>= 4.0$ GeV}, Phys.Rev. D {\bf 92}, 092008 (2015).
 
\bibitem{MINERvA-1pi_1} C.L.~McGivern {\it et al.} (MINER$\nu$A Collaboration), {\it Cross sections for $\nu_\mu$ and $\bar \nu_\mu$ induced pion production on hydrocarbon in the few-GeV region using MINER$\nu$A} , Phys. Rev. D {\bf 94}, 052005 (2016).

\bibitem{MINERvA-1pi_2} P.\.Stowell {\it et al.} (MINER$\nu$A Collaboration), {\it Tuning the GENIE Pion Production Model with MINER$\nu$A Data }, arXiv:1903.01558 [hep-ex] (2019)

\bibitem{T2K1pi_water} K.~Abe {\it et al.} (T2K Collaboration), {\it First measurement of the muon neutrino charged current single pion production cross section on water with the T2K near detector}', Phys. Rev. D{\bf 95} 012010 (2017).

\bibitem{Betancourt:2018bpu}
  M.~Betancourt {\it et al.},
 {\it Comparisons and Challenges of Modern Neutrino Scattering Experiments (TENSIONS2016 Report)},
  Phys.\ Rept.\  {\bf 773-774} (2018) 1.

\bibitem{Sobczyk:2014xza}
  J.~T.~Sobczyk and J.~Żmuda,
  {\it Investigation of recent weak single-pion production data},
  Phys.\ Rev.\ C {\bf 91} (2015) no.4,  045501

\bibitem{Cai:2019jzk}
  T.~Cai, X.~Lu and D.~Ruterbories,
  {\it Pion-Proton Correlation in Neutrino Interactions on Nuclei},
  arXiv:1907.11212 [hep-ex].

\bibitem{Martini:2009} M.~Martini, M.~Ericson, G.~Chanfray, J.~Marteau, {\it Unified approach for nucleon knock-out and coherent and incoherent pion production in neutrino interactions with nuclei}, Phys. Rev. C {\bf80}, 065501(2009).

\bibitem{rein-sehgal} D.~Rein, L.M.~Sehgal, {\it Neutrino-excitation of baryon resonances and single pion production}, Annals of Physics 133 (1981) http://dx.doi.org/10.1016/0003-4916(81)90242-6.

\bibitem{Salcedo:1987md} L.L.~Salcedo, E.~Oset, M.J.~Vicente-Vacas, C.~Garcia-Recio, {\it Computer Simulation of Inclusive Pion Nuclear Reactions}, Nucl. Phys., A {\bf484}, (1988).

\bibitem{Hernandez:2007qq} E.~Hernandez, J.~Nieves, M.~Valverde, {\it Weak Pion Production off the Nucleon},  Phys. Rev. D {\bf 76} 033005 (2007).

\bibitem{Buss:2007ar} O.~Buss, T.~Leitner, U.~Mosel, L.~Alvarez-Ruso {\it The Influence of the nuclear medium on inclusive electron and neutrino scattering off nuclei},  Phys. Rev. C {\bf 76} 035502 (2007).

\bibitem{Praet:2008yn} C.~Praet, O.~Lalakulich, N.~Jacowicz, J.~Rychebusch, 
{\it Delta-mediated pion production in nuclei}, Phys. Rev. C {\bf 79} 044603 (2009).
   
\bibitem{Serot:2012rd} B.D.~ Serot, X.~ Zhang, {\it Neutrino production of Photons and Pions From Nucleons in a Chiral Effective Field Theory for Nuclei}, Phys. Rev. C {\bf 86} 015501 (2012). 

\bibitem{Ivanov:2015aya} M.V.~Ivanov, G.D.~Megias, R.~Gonz\'{a}lez-Jim\'{e}nez,
     0.~Moreno, M.B.~Barbaro, J.A.~Caballero, T.W.~Donnelly, {\it Charged-current inclusive neutrino cross sections in the SuperScaling model including quasielastic, pion production and meson-exchange contributions}, J. Phys. G {\bf 43} 4 0451001 (2016).
     
\bibitem{Alam:2015gaa} M.~Rafi Alam, M.~Sajjad Athar, S.~Chaulan, S.K.~Singh, 
      {\it Weak charged and neutral current induced one pion
                        production off the nucleon},  Int. J. Mod. Phys. E {\bf 25} 02 01650010 (2016).
                        
\bibitem{Nakamura:2015rta} S.X.~Nakamura, H.~Kamano, T.~Sato, {\it Dynamical coupled-channels model for neutrino-induced meson productions in resonance region}, Phys. Rev. D {\bf 92} 074024 (2015). 
                        
\bibitem{Hernandez:2013jka} E.~Hern\'{a}ndez, J.~Nieves, M.J.~Vicente Vacas,
{\it Single $\pi$ production in neutrino-nucleus scattering}, Phys. Rev. D {\bf 87} 113009 (2013).

\bibitem{Gonzalez-Jimenez:2016qqq} R.~Gonz\'{a}lez-Jim\'{e}nez, N.~Jacowicz, K.~Niewczas, J.~Nys, V.~Pandey, T.~Van Cuyck, N.~Van Dessel, {\it Electroweak single-pion production off the nucleon: from threshold to high invariant masses}, Phys. Rev. D {\bf 95} 113007 (2017).

\bibitem{Sobczyk.:2012zj} J.T.~Sobczyk, J.~Zmuda, {\it Impact of nuclear effects on weak pion production at energies below 1 GeV}, Phys. Rev. C{\bf 87} 065503 (2013).

\bibitem{Mosel:2017nzk} U.~Mosel, K.~Gallmeister, {\it Muon-neutrino-induced charged current pion production on nuclei}, Phys.Rev. C {\bf 96} 015503 (2017).  Addendum:  Phys. Rev. C {\bf 99}, 035502  (2019)

\bibitem{Kabirnezhad:2017jmf} M.~Kabirnezhad, {\it Single pion production in neutrino-nucleon Interactions} Phys.\ Rev.\ D {\bf 97} 013002 (2018).

\bibitem{Sobczyk:2018ghy}   J.E.~Sobczyk, E.~Hernández, S.X.~Nakamura, J.~Nieves and T.~Sato,
  {\it Angular distributions in electroweak pion production off nucleons: odd parity hadron terms, strong relative phases and model dependence},
  Phys.\ Rev.\ D {\bf 98}, no. 7, 073001 (2018).
  
\bibitem{t2k-nim} K.~Abe {\it et al.} (T2K Collaboration), Nucl. Instrum. Meth. A {\bf659}, 106 (2011).

\bibitem{PhysRevD.87.012001} K.~Abe {\it et al.} (T2K Collaboration), {\it T2K neutrino flux prediction}, Phys. Rev. D {\bf 87}, 012001 (2013); ; erratum Phys. Rev. D {\bf87}, 019902 (2013).

\bibitem{t2k-p0d} S.~Assylbekov {\it et al.}, {\it The T2K ND280 Off-Axis Pi-Zero Detector}, Nucl. Instrum.  Meth. A {\bf686},  48 (2012).

\bibitem{t2k-ecal} D.~Allan {\it et al.}, {\it The Electromagnetic Calorimeter for the T2K Near Detector ND280}, JINST {\bf8} P10019 (2013)

\bibitem{Aoki:2012mf} S.~Aoki {\it et al.}, {\it The T2K Side Muon Range Detector (SMRD)},  Nucl.Instrum.Meth. A{\bf 698} (2013) 135

\bibitem{t2k-tpc} N.~Abgrall {\it et al.},  {\it Time projection chambers for the T2K near detectors}, Nucl. Instrum. Meth. A {\bf637}, 25 (2011)

\bibitem{t2k-fgd} P.~Amaudruz, {\it et al.}, {\it The T2K Fine-Grained detectors}, Nucl. Instrum. Meth. A {\bf696}, 1 (2012).\

\bibitem{Sanchez:2015yvw} F.~S\'{a}nchez, {\it Possibility of measuring Adler angles in charged current single pion neutrino-nucleus interactions}, Phys.\ Rev.\ D {\bf 93}, no. 9, 093015 (2016).

\bibitem{d'Agostini-unfolding} G.~D`Agostini, {\it A multidimensional
    unfolding method based on Bayes’ theorem}, Meth. in Phys. Res. A
  {\bf362} (1995) 487.

 \bibitem{Ferrari:2005zk} A.~Ferrari, P.R.~Sala, A.~Fasso, J.~Ranft, {\it FLUKA: A multi-particle transport code}, CERN-2005-10 (2005), INFN/TC\_05/11, SLAC-R-773.

 \bibitem{Battistoni:2007zzb} G.~Battistoni, S.~Muraro, P.R.~Sala, F.~Cerutti, A.~Ferrari {\it et al.}, {\it The FLUKA code: Description and benchmarking}, AIP Conf.Proc., 896, (2007). Note: For this work we used FLUKA2008, which was the most recent version at the time of this study. A new version, FLUKA2011, has been released and the comparison with data would be different.
  
 \bibitem{Abgrall:2015hmv} N.~Abgrall {\it et al.} (NA61/SHINE Collaboration), {\it Measurements of $\pi^\pm$, $K^\pm$, $K^0_S$, $\Lambda$ and proton production in proton-carbon interactions at 31 GeV/$c$ with the NA61/SHINE spectrometer at the CERN SPS}, Eur.\ Phys.\ J.\ C {\bf 76}, no. 2, 84 (2016).

 \bibitem{PhysRevC.85.035210} N.~Abgrall {\it et al.} (NA61/SHINE Collaboration), {\it Measurement of production properties of positively charged kaons in proton-carbon interactions at 31 GeV/$c$}, Phys. Rev. C {\bf85}, 035210 (2012).
 
\bibitem{PhysRevC.84.034604} N.~Abgrall {\it et al.} (NA61/SHINE  Collaboration), {\it Measurements of cross sections and charged pion spectra in proton-carbon interactions at 31 GeV/$c$},  Phys. Rev. C {\bf 84}, 034604 (2011).
 
\bibitem{eichten} T.~Eichten {\it et al.}, {\it Particle production in proton interactions in nuclei at 24 GeV/c}, Nucl. Phys. B {\bf44} (1972).
   
\bibitem{allaby} J.V.~Allaby, {\it et al.}, {\it High-energy particle spectra from proton interactions at 19.2 GeV/c}, CERN 70-12, (1970).
   
 \bibitem{PhysRevC.77.015209} I.~Chemakin, V.~Cianciolo, B.A.~Cole, R.C.~Fernow, A.D.~Frawley,M.~Gilkes, S.~Gushue, E.P.~Hartouni,
H.~Hiejima {\it et al.} (E910 Collaboration), {\it Pion production by protons on a thin beryllium target at 6.4, 12.3, and 17.5 GeV/$c$ incident proton momenta}, Phys. Rev. C {\bf77}, 015209 (2008); erratum Phys. Rev. C {\bf77}, 049903(E) (2008).

\bibitem{GEANT3} R.~Brun, F.~Carminati, S.~Giani, CERN-W5013 (1994).
    
\bibitem{GCALOR} C.~Zeitnitz and T.A.~Gabriel, {\it Proc. of International Conference on Calorimetry in High Energy Physics}, (1993).

\bibitem{Hayato:2002sd} Y.~Hayato, {\it NEUT}, Nucl.Phys.Proc.Suppl. 112,10.1016/S0920-5632(02)01759-0 (2002).

\bibitem{Hayato:2009} Y.~Hayato, {\it A neutrino interaction simulation program library NEUT}, Acta Phys.Polon. B{\bf40} (2009).
  
\bibitem{llewellyn-smith} Y.~Hayato, {\it Neutrino Reactions at Accelerator Energies}, Phys. Rep. 3 (1972).

\bibitem{smith-moniz} R.A.~Smith and E.J.~Moniz, {\it Neutrino Reactions on Nuclear Targets}, Nucl.Phys. B{\bf43} (1972); erratum: Nucl.Phys. B{\bf101} (1975) 547.

\bibitem{rein-sehgal-coha} D.~Rein, L.~Sehgal, {\it Coherent $\pi^\circ$ production in neutrino reactions}, Nucl. Phys. B {\bf223} (1983) 29

\bibitem{rein-sehgal-cohb} D.~Rein and L.~Sehgal, {\it PCAC and the deficit of forward muons in $\pi^+$ production by neutrinos}, Phys. Lett. B {\bf657} (2007) 207

\bibitem{Gluck:1998xa} M.~Gluck, E.~Reya, and A.~Vogt, {\it Dynamical parton distributions revisited}, Eur. Phys. J. C{\bf5} (1998), hep-ph/9806404.\  

\bibitem{Bodek:2003wd} A.~Bodek, U.K.~Yang, {\it Modeling neutrino and electron scattering inelastic cross sections in the few GeV region with effective LO PDFs}, AIP Conf. Proc. {\bf670}, (2003) 10.1063/1.1594324; http://dx.doi.org/10.1063/1.1594324.\  

\bibitem{Andreopoulos:2009rq} C.~Andreopoulos, A.~Bell, D.~Bhattacharya, F.~Cavanna, J.~Dobson, {\it et al.},{\it The GENIE Neutrino Monte Carlo Generator}, Nucl. Instrum. Methods Phys. Res., Sect. A{\bf614} (2010).  

\bibitem{GEANT4} S.~Agostinelli et al. (GEANT4), Nucl. Instrum. Meth. {\bf A} 506, 250 (2003).

\bibitem{Kuzmin:2007kr} K.S.~Kuzmin, V.V.~Lyubushkin,  V.A.~Naumov, {\it Quasielastic axial-vector mass from experiments on neutrino-nucleus scattering}, Eur. Phys. J. C{\bf54} (2008).    
   
\bibitem{Kuzmin:2006dh} K.S.~Kuzmin, V.V.~Lyubushkin, and V.A.~Naumov, Acta Phys. Polon. B {\bf37} 2337 (2006).

\bibitem{Abe:2015awa} K.~Abe, {\it et al.} (T2K Collaboration), {\it Measurements of neutrino oscillation in appearance and disappearance channels by the T2K experiment with $6.6\times10^{20}$ protons on target}, Phys. Rev. D {\bf 91}, 072010 (2015).

\bibitem{Benhar:1995} O.~Benhar, B.G.~Zakharov, N.N.~Nikolaev, S.~Fantoni,{\it Nuclear Effects in the Diffractive Electroproduction of s Anti-s Mesons}, Phys. Rev. Lett. 74 (1995). 



\bibitem{Minervapi0} O.~Altinok {\it et al.} (MINER$\nu$A Collaboration), {\it Measurement of $\nu_{\mu}$ charged-current single $\pi^{0}$ production on hydrocarbon in the few-GeV region using MINER$\nu$A} , Phys. Rev. D {\bf 96}, 072003 (2017).

\end{thebibliography}
\end{document}